\documentclass[twocolumn]{revtex4}

\usepackage{ragged2e}
\usepackage{amsmath}
\usepackage{graphicx}
\usepackage{rotating}

\begin{document}

\title{A systematic analysis of transverse momentum spectra of $J/\psi$
mesons \\ in high energy collisions \vspace{0.5cm}}

\author{Xu-Hong Zhang$^{1,}$\footnote{xhzhang618@163.com; zhang-xuhong@qq.com}}

\author{Fu-Hu Liu$^{1,}$\footnote{Corresponding author: fuhuliu@163.com; fuhuliu@sxu.edu.cn}}

\author{Khusniddin K. Olimov$^{2,}$\footnote{Corresponding author: khkolimov@gmail.com}}

\affiliation{$^1$Institute of Theoretical Physics \& Collaborative
Innovation Center of Extreme Optics \& State Key Laboratory of \\
Quantum Optics and Quantum Optics Devices, Shanxi University,
Taiyuan 030006, China
\\
$^2$Laboratory of High Energy Physics, Physical-Technical
Institute of SPA ``Physics-Sun" of Uzbek Academy of Sciences, \\
Chingiz Aytmatov str. $2^b$, Tashkent 100084, Uzbekistan}

\begin{abstract}

\vspace{0.5cm}

\noindent {\bf Abstract:} We aggregate the transverse momentum
spectra of $J/\psi$ mesons produced in high energy gold-gold
(Au-Au), deuteron-gold ($d$-Au), lead-lead (Pb-Pb), proton-lead
($p$-Pb), and proton-(anti)proton ($p$-$p(\overline{p})$)
collisions measured by several collaborations at the Relativistic
Heavy Ion collider (RHIC), the Tevatron Proton-Antiproton
Collider, and the Large Hadron Collider (LHC). The collision
energy (the center-of-mass energy) gets involved in a large range
from dozens of GeV to 13 TeV (the top LHC energy). We consider two
participant or contributor partons, a charm quark and an
anti-charm quark, in the production of $J/\psi$. The probability
density of each quark is described by means of the modified
Tsallis--Pareto-type function (the TP-like function) while
considering that both quarks make suitable contributions to the
$J/\psi$ transverse momentum spectrum. Therefore, the convolution
of two TP-like functions is applied to represent the $J/\psi$
spectrum. We adopt the mentioned convolution function to fit the
experimental data and find out the trends of the power exponent,
effective temperature, and of the revised index with changing the
centrality, rapidity, and collision energy. Beyond that, we
capture the characteristic of $J/\psi$ spectrum, which is of great
significance to better understand the production mechanism of
$J/\psi$ in high energy collisions.
\\
\\
{\bf Keywords:} Transverse momentum spectrum, $J/\psi$ meson,
TP-like function, convolution
\\
\\
{\bf PACS:} 12.40.Ee, 13.85.Hd, 24.10.Pa
\\
\\
\end{abstract}

\maketitle

\section{Introduction}

Quantum chromodynamics (QCD) is the standard dynamics theory and
an important part of the standard model, which is applicable in
the study of heavy quark pair production and
correlation~\cite{1,2,3,4,5,6} such as the transverse momentum
spectra, nuclear modification factor, azimuthal correlation,
anisotropic flow, and so on. QCD is a kind of non-Abelian gauge
field theory~\cite{6a}, which implies that the strong interactions
between quarks have three basic characteristics. Firstly, it can
explain the asymptotic freedom characteristics proposed in the
inelastic electron-proton and electron-deuteron
scattering~\cite{6b}. Secondly, it can explain the color
confinement which shows quarks and anti-quarks cannot be separated
due to very strong interactions. Lastly, it can explain the
spontaneous break of the symmetry of the chirality. Understanding
these characteristics is of great necessity for researchers to
study the interactions among particles and their mechanisms of
evolution, structure, and decay~\cite{6c,6d,6e,6f,6g}.

The heavy flavor quarkonium is a bound state formed by the heavy
flavor quark and anti-quark~\cite{7}. It plays an important role
in the theoretical research of QCD. In hadron induced high energy
collisions, the generation of heavy quarkonium can be divided into
two processes: One is the appearance of heavy quark pairs and the
other one is the evolution of heavy quark pairs into hadrons. The
former process can be calculated and analyzed by the perturbative
QCD theory~\cite{9a,9b,9c}. Particularly, due to the peculiar
reason of the heavy quarkonium, in which the relativistic effect
can be neglected, some special theories such as the
non-relativistic QCD theory~\cite{9d,9e,9f} can be used to
calculate and analyze the production process.

As the basic theory of strong interactions of
particles~\cite{1,2,3,4,5,6}, QCD predicts that the hadronic
matter can be heated to a very high temperature when it
experiences very strong interactions. Then, the system will go
through a phase transition from the hadron matter to quark-gluon
plasma (QGP) in the process~\cite{10,11,12}. The experiment of
relativistic heavy ion collisions is the only way of achieving a
QCD phase transition in laboratory conditions~\cite{12a,12b,12c}.
Nevertheless, the lifetime of the produced QGP experiencing this
phase transition can only reach the order of 10 fm/$c$ (from a few
to dozens of fm/$c$)~\cite{12d,12e}, which cannot be directly
observed in experiments. To detect QGP and study its properties,
one has to use an indirect method. For example, one may study the
spectrum properties of heavy quarkoniums to obtain the excitation
degree (temperature) of emission source which is related to the
information on QGP.

$J/\psi$ meson is the bound state of charm and anti-charm
($c\overline{c}$) quarks, where the constituent mass of charm
quark is about 1.6 GeV/$c^2$~\cite{7}. As the first heavy
quarkonium discovered experimentally, it has been extensively
studied in high energy collisions. In addition, the constraint of
$J/\psi$ is considered as an important signal for the generation
of QGP~\cite{13,14,15,15a,15b}. The yield of $J/\psi$ in
electron-positron collisions is higher than that in nuclear
collisions~\cite{15c,15d}, so the decay of $J/\psi$ in nuclear
collisions is an ideal way and medium for studying the hadron
spectrum and finding new particles. $\it \Upsilon$ meson is a
bound state of bottom and anti-bottom ($b\overline{b}$) quarks,
where the constituent mass of bottom quark is about 4.6
GeV/$c^2$~\cite{7}. The masses of both $J/\psi$ and $\it \Upsilon$
are very large, which leads to the change scale of energy
(momentum) in the collision process to the order of GeV (GeV/$c$)
when we study their structural properties. In addition,
$c+\overline{c}$ and $b+\overline{b}$ can form new $J/\psi$ and
$\it \Upsilon$ respectively, which can be researched by
non-relativistic approach~\cite{9d,9e,9f}.

We note that the theories and models based on QCD and related idea
are complex in the calculation process~\cite{9a,9b,9c,9d,9e,9f}.
The complex calculation limits the applications of these theories
and models in comparison with experimental data. We hope that we
could use a simple idea and formalism to describe uniformly the
spectra of various particles, in particular the spectra of heavy
quarkoniums such as $J/\psi$ due to its abundant early production
in the collisions and wide transverse momentum distribution. In
the framework of the multi-source thermal
model~\cite{16,17,18,18a}, we have used the idea of quark
composition describing tentatively and uniformly the spectra of
various particles in our recent work~\cite{64,64a}. It is
interesting for us to test further the idea systematically, but
the idea of two contributor quarks or partons is used due to the
fact that some particles such as leptons have no quark
composition. Of course, the quark composition and two contributor
quarks for heavy quarkoniums are the same.

In this paper, we test systematically the idea of two contributor
partons which contribute to the transverse momentum spectrum of
$J/\psi$. The experimental data are collected from gold-gold
(Au-Au)~\cite{19,20}, deuteron-gold ($d$-Au)~\cite{21}, lead-lead
(Pb-Pb)~\cite{22,23}, proton-lead ($p$-Pb)~\cite{24,25,26,27}, and
proton-(anti)proton ($p$-$p(\overline{p})$)
collisions~\cite{28,29,30,31,32,33,34,35,36,37,38} over an energy
range from dozens of GeV at the Relativistic Heavy Ion Collider
(RHIC) to 13 TeV at the Larger Hadron Collider (LHC). Among the
RHIC and LHC, there is the Tevatron Proton-Antiproton Collider
from which we cited the data in $p$-$\overline{p}$ collisions at
1.96 TeV. These studies are useful for us to understand one of the
three basic characteristics of QCD, the color confinement due to
the very strong interactions among quarks and anti-quarks.

\section{Formalism and method}

According to the multi-source thermal model~\cite{16,17,18,18a},
we may think that a few emission sources are formed in high energy
collisions. For nuclear fragments from the projectile and target
in nucleus-nucleus collisions, the sources can be nucleons and
nucleon clusters. For produced particles such as pions, kaons, and
$J/\psi$, the sources can be participant or contributor quarks or
gluons, though the contributors $c+\overline{c}$ may be from gluon
fusion at the first. The properties of sources can be described by
different statistics such as the Boltzmann-Gibbs, Fermi-Dirac,
Bose-Einstein, and Tsallis statistics. There are some relations
among these statistics due to the fact that they may result in
similar or different distributions while describing the spectra of
particles.

The Tsallis distribution describes the transverse momentum
($p_{\rm T}$) spectra in wider range than the Boltzmann-Gibbs
distribution, though the former is derived from the
latter~\cite{39,40,41}. Also, the latter is a special case of the
former in which the entropy index $q=1$. Indeed, the former is
widely used in high energy collisions from a few GeV to 13 TeV
(the top LHC energy) to parameterize the $p_{\rm T}$ spectrum of
final-state particles, which justifies its usage in the present
work. The form of the Tsallis distribution is expressed
as~\cite{42,43,44,45,46,47,48}
\begin{align}
\begin{split}
&E\frac{d^{3}N}{d^{3}p}=\frac{1}{2\pi p_{\rm T}}\frac{d^{2}N}{dp_{\rm T}dy}\\
&=\frac{dN}{dy}\frac{(n-1)(n-2)}{2\pi nT
[nT+m_{0}(n-2)]}\bigg(1+\frac{m_{\rm T}-m_{0}}{nT}\bigg)^{-n}.
\end{split}
\end{align}
Here, $E$, $p$, $N$, $y$, $m_0$, $m_{\rm T}$, $n$, and $T$ denote
the energy, momentum, particle number, rapidity, rest mass,
transverse mass, power exponent, and effective temperature,
respectively. The transverse mass is given by $m_{\rm
T}=\sqrt{p_{\rm T}^{2}+m_0^2}$~\cite{49,50,51,52,53,54}. In
particular, $n=1/(q-1)$, and the entropy index $q$ describes the
degree of equilibrium. The closer the parameter $q$ to 1, the more
equilibrated the emission source is.

According to the form of Tsallis distribution Eq. (1), as $p_{\rm
T}\gg m_0$, $m_0$ can be ignored, followed by $Ed^{3}N/dp^{3}
\propto p_{\rm T}^{-n}$. Then it can be observed that the
particles are distributed in accordance with the inverse power
law. This is the distribution type of particles produced by the
hard scattering process in the high energy collision
process~\cite{55,56,57,58} and in high $p_{\rm T}$ region. In the
non-relativistic limit $(p_{\rm T}\ll m_0)$ condition, there is
$m_{\rm T}-m_{0}=p_{\rm T}^{2}/2m_0=E_{\rm T}^{\rm classical}$,
showing $Ed^{3}N/dp^{3} \propto e^{-E_{\rm T}^{\rm classical}/T}$,
where $E_{\rm T}^{\rm classical}$ is the transverse energy in the
non-relativistic limit. We call this distribution the
thermodynamic statistical distribution, that is the Boltzmann
distribution. Here, we have only discussed the two special cases
($p_{\rm T}\gg m_0$ and $p_{\rm T}\ll m_0$), though they are not
used by us in the present work.

Usually, the empirical formula, the Tsallis--Pareto-type function,
is adopted to outline the $p_{\rm T}$
spectrum~\cite{59,60,61,62,63}. The general form of the mentioned
function is
\begin{align}
f(p_{\rm T})=C \times p_{\rm T} \times \bigg(1+\frac{m_{\rm
T}-m_{0}}{nT}\bigg)^{-n}
\end{align}
which is equivalent to Eq. (1) in the form of probability density
function, where $C$ is the parameter dependent normalization
constant. As the probability density function, Eq. (2) is
normalized as $\int_{0}^{\infty} f(p_{\rm T})dp_{\rm T}=1$. In Eq.
(2), $n$ and $T$ reflect the degrees of non-equilibrium and
excitation of the source respectively. Larger $n$ corresponds to
more equilibrium, and larger $T$ corresponds to higher excitation.

In the lower $p_{\rm T}$ range, due to the contribution of light
flavor resonance decay, Eq. (2) cannot describe the spectra of
light particles very well. For $J/\psi$, the feed-down
contribution is more complicated, which is minimal at low $p_{\rm
T}$ and grows with growing $p_{\rm T}$. This renders that Eq. (2)
also fails to describe the $J/\psi$ spectra. As a result, we ought
to empirically add a revised index $a_{0}$ on $p_{\rm T}$ to
modify Eq. (2). Then the revised Eq. (2) becomes~\cite{64,64a,65}
\begin{align}
f(p_{\rm T}) = C \times p_{\rm T}^{a_{0}} \times
\bigg(1+\frac{m_{\rm T}-m_{0}}{nT}\bigg)^{-n}.
\end{align}
Both the normalization constants $C$ in Eqs. (2) and (3) are
different, though we have used the same symbol. The two constants
are also the parameter dependent. Compared to Eq. (2), Eq. (3) can
be used to describe the spectrum in the entire transverse momentum
range, having a broader application. For purpose of convenience,
as in refs.~\cite{64,64a,65}, we also call Eq. (3) the TP-like
function in this work. In the TP-like function, the meanings of
$n$ and $T$ remain unchanged as what they are in Eq. (2), though
their values may be changed.

The discovery of $J/\psi$ provides a direct evidence for the
existence of charm quarks, which makes the study of hadron
structure theory presenting a new situation. We may think that in
the formation of $J/\psi$ there are two participant or contributor
(anti-)charm quarks taking part in the collisions. Let $p_{t1}$
and $p_{t2}$ denote the contributions of quarks 1 and 2 to the
transverse momentum of $J/\psi$ respectively. The probability
density function $f_{1}(p_{t1})$ ($f_{2}(p_{t2})$) obeyed by
$p_{t1}$ ($p_{t2}$) is assumed to be Eq. (3). We have
\begin{align}
f_{1}(p_{t1})=C_1p_{t1}^{a_{0}}
\bigg(1+\frac{\sqrt{p_{t1}^2+m_{1}^2}-m_{1}}{nT}\bigg)^{-n},
\end{align}
\begin{align}
f_{2}(p_{t2})=C_1p_{t2}^{a_{0}}
\bigg(1+\frac{\sqrt{p_{t2}^2+m_{2}^2}-m_{2}}{nT}\bigg)^{-n}.
\end{align}
Here, the two normalization constants $C_1$ and $C_2$ are
parameter dependent. $m_1$ and $m_2$ are the constituent masses of
quarks 1 and 2 respectively, both are 1.6 GeV/$c^2$ for charm and
anti-charm quarks~\cite{7} used in this work. Because of the two
quarks taking part in the same collisions, the parameters $n$,
$T$, and $a_0$ in Eqs. (4) and (5) are separately the same.

The transverse momentum distribution of $J/\psi$ is given by the
convolution of two TP-like functions~\cite{64,64a,65}. We have
\begin{align}
\begin{split}
f(p_{\rm T})&=\int_{0}^{p_{\rm T}} f_{1}(p_{t1})f_{2}(p_{\rm T}-p_{t1})dp_{t1}\\
&=\int_{0}^{p_{\rm T}} f_{2}(p_{t2})f_{1}(p_{\rm
T}-p_{t2})dp_{t2},
\end{split}
\end{align}
where the functions $f_1(x)$ and $f_2(x)$, as well as the
parameters $n$, $T$, and $a_0$, are given in Eqs. (4) and (5). It
should be noted that we have used two contributors. The
convolution of two-parton contributions is applicable even for the
spectra of particles with complex quark composition. For example,
we may use the convolution of two-parton contributions fitting the
spectra of various jets~\cite{65}. As the probability density
function, Eq. (6) is applicable in high energy collisions with
small or large system, no matter what the density of produced
particles is. In addition, Eq. (6) results in similar curve as
Eqs. (4) and (5) with different parameters~\cite{64,64a}, though
the form of Eq. (6) is more complex due to the convolution.

In the above discussions, we assume that $p_{\rm
T}=p_{t1}+p_{t2}$, which is operated to describe the relationship
among $p_{\rm T}$ of $J/\psi$, $p_{t1}$ and $p_{t2}$ contributed
by quarks 1 and 2, respectively. This treatment assumes the
azimuth angle $\phi_1$ of vector $\vec{p}_{t1}$ being equal to the
azimuth angle $\phi_2$ of vector $\vec{p}_{t2}$. More
generally~\cite{64,64a,66}, if $\phi_1\neq\phi_2$, we have $p_{\rm
T}=\sqrt{p_{t1}^{2}+p_{t2}^{2}+
2p_{t1}p_{t2}\cos|\phi_{1}-\phi_{2}|}$. In particular, if
$\vec{p}_{t1}$ is perpendicular to $\vec{p}_{t2}$, i.e.
$|\phi_1-\phi_2|=\pi/2$, we have $p_{\rm
T}=\sqrt{p_{t1}^{2}+p_{t2}^{2}}$. If $\vec{p}_{t1}$ is opposite to
$\vec{p}_{t2}$, i.e. $|\phi_1-\phi_2|=\pi$, we have $p_{\rm
T}=|p_{t1}-p_{t2}|$. Our explorations show that the relationship
of $p_{\rm T}=p_{t1}+p_{t2}$ due to $\phi_1=\phi_2$, i.e. Eq. (6)
based on Eqs. (4) and (5), is more easy to fit the data.

If we assume other relationships, Eq. (6) is not applicable. Thus,
we need to explore new form of Eq. (3), which is also the topic
for us~\cite{66,67}. If the analytical expression of Eq. (6) is
not available for other relationships, the Monte Carlo method can
be used to obtain $p_{t1}$, $p_{t2}$, and $p_{\rm T}$. The $p_{\rm
T}$ distribution is then obtained from the statistics. For
example, for the more general case of $\phi_1\neq\phi_2$, we have
the expression of $p_{\rm
T}=\sqrt{(p_{t1}\cos\phi_1+p_{t2}\cos\phi_2)^2
+(p_{t1}\sin\phi_1+p_{t2}\sin\phi_2)^2}$. This expression can be
extended to the case of three or more contributor partons if we
add the third or more items in the components. For the special
case of $|\phi_1-\phi_2|=\pi/2$, the analytical expression of Eq.
(6) is changed~\cite{66,67}, and the form of Eq. (3) is also
changed~\cite{67}. For the special case of $|\phi_1-\phi_2|=\pi$,
the analytical expression of Eq. (6) is applicable, though the
form of Eq. (3) is not applicable. In the calculation, the
conservations of energy and momentum should be satisfied
naturally.

It is understandable that we use $p_{\rm T}=p_{t1}+p_{t2}$ in the
present work. From the point of view of energy, we have the
relationship among the energy $E$ of $J/\psi$, $E=E_1+E_2$, where
$E_{1}$ and $E_{2}$ are energies contributed by quarks 1 and 2,
respectively. In term of transverse mass and rapidity, we have the
relationship $m_T\cosh y =m_{t1}\cosh y_1 +m_{t2}\cosh y_2$, where
$m_{t1} =\sqrt{p_{t1}^2+m_{01}^2}$, $m_{t2}
=\sqrt{p_{t2}^2+m_{02}^2}$, $m_{01}$ denotes the mass of quark 1,
$m_{02}$ denotes the mass of quark 2, $y_1$ denotes the rapidity
of quark 1, and $y_2$ denotes the rapidity of quark 2. In the
given narrow rapidity range and neglecting the mass, we have
approximately $p_{\rm T}=p_{t1}+p_{t2}$.

Although the case of $\phi_1=\phi_2$ is a special one, the
relationship of $p_{\rm T}=p_{t1}+p_{t2}$ and the convolution of
two TP-like functions can fit easily the spectra of various
particles and jets~\cite{64,64a,65}. From our point of view, we do
not think that the more general case of $\phi_1\neq\phi_2$ and the
special cases of $|\phi_1-\phi_2|=\pi/2$ and $|\phi_1-\phi_2|=\pi$
are more easy to fit the spectra due to the calculation itself,
though the idea is practicable. In particular, from the point of
view of the contributor partons, but not the constituent quarks,
the spectra of leptons and jets can be easily fitted from the
relationship of $p_{\rm T}=p_{t1}+p_{t2}$ and the convolution of
two TP-like functions~\cite{64,64a,65}. This confirms the validity
of Eq. (6) based on two contributor sources in the framework of
multi-source thermal model~\cite{16,17,18,18a}.

We would like to emphasize here that we have used two contributor
partons as the projectile and target particles/nuclei, no matter
what the final-state products are~\cite{64,64a,65}. For $J/\psi$,
the two contributor partons (charm and anti-charm quarks) and the
constituent quarks are coincidentally equal to each other. For
baryons, the two contributor partons are not equal to the three
constituent quarks. For jets, the two contributor partons are not
equal to the sets of two or three constituent quarks, too. For
leptons, the two contributor partons have no corresponding
constituent quarks. We may consider that two light or slow
contributor partons produce leptons and baryons, while two heavy
or fast contributor partons produce jets, no matter what the
structures of leptons, baryons, and jets are. In fact, the two
contributor partons are regarded as two energy resources, but not
the constituents.

\section{Results and discussion}

\subsection{Comparison with data}

Figure 1 shows the transverse momentum spectra, $Bd^{2}N/(2\pi
p_{\rm T} dp_{\rm T}dy)$, of $J/\psi$ produced in Au-Au collisions
at center-of-mass energy per nucleon pair $\sqrt{s_{\rm NN}}=$ (a)
39, (b) 62.4, and (c--e) 200 GeV, where $B$ denotes the branching
ratio. The symbols in panels (a--c) represent the experimental
data of $p_{\rm T}$ spectra measured by the STAR
Collaboration~\cite{19} in the mid-rapidity interval of $|y|<1$
and in the centrality class of 0--60$\%$ and its subclasses of
0--20$\%$, 20--40$\%$, and 40--60$\%$. The symbols in panels (d)
and (e) represent the experimental data of $p_{\rm T}$ spectra
measured by the PHENIX Collaboration~\cite{20} in the rapidity
intervals of (d) $|y|<0.35$ and (e) $y\in[1.2,2.2]$ and in the
centrality classes of 0--20$\%$, 20--40$\%$, 40--60$\%$, and
60--92$\%$. Some data sets are scaled by multiplying or dividing
different values marked in the panels for clear indication. The
solid curves are our fitting results by using the convolution of
two TP-like functions, i.e. Eq. (6) based on Eqs. (4) and (5). For
comparison, the dot-dashed curves are our results refitted by Eq.
(6) with $a_0=1$. The values of fitting parameters $n$, $T$, and
$a_0$ are listed in Table 1 with the normalization constant $N_0$,
$\chi^2$, and the number of degree of freedom (ndof). We use
$\chi^{2}$ to characterize the fitting deviation between the
experimental data and our fit function and curve. For the given
data and fit function, the smaller $\chi^{2}$, the better the
fitting result, and the closer to the experimental results. If
$\chi^{2}<1$, it is rounded to 1 or a decimal fraction; Otherwise,
it is rounded to an integer. In the case of ndof is less than or
equal to 0, we use $``-"$ to mention in the table. One can see
that the mentioned function with changeable $a_0$ fits
satisfactorily the experimental data in Au-Au collisions measured
by the STAR and PHENIX Collaborations at the RHIC, though in many
cases the fits with $a_0=1$ are comparable to those with
changeable $a_0$.

Similar to Figure 1, Figure 2 shows the transverse momentum
spectra, $Bd^{2}N/(2\pi p_{\rm T}dp_{\rm T}dy)$, of $J/\psi$
produced in $d$-Au collisions at 200 GeV. The symbols represent
the experimental data measured by the PHENIX
Collaboration~\cite{21}. Panels (a) and (b) show the spectra of
$J/\psi\rightarrow\mu^{+}\mu^{-}$ at the backward rapidity of
$-2.2<y<-1.2$ and the forward rapidity of $1.2<y<2.2$
respectively, with the centrality classes of 0--20$\%$,
20--40$\%$, 40--60$\%$, and 60--88$\%$. Panel (c) shows the
spectra of $J/\psi\rightarrow e^{+}e^{-}$ at mid-rapidity
$|y|<0.35$ with the same centrality classes as panels (a) and (b).
Panel (d) shows the spectra in minimum-bias collisions with
rapidity intervals of $-2.2<y<-1.2$, $1.2<y<2.2$, and $|y|<0.35$.
The solid curves are our fitting results by Eq. (6), and the
dot-dashed curves are our results refitted by Eq. (6) with
$a_0=1$. The values of fitting parameters are listed in Table 2
with $N_0$, $\chi^2$, and ndof. One can see that the mentioned
function with changeable $a_0$ fits satisfactorily the
experimental data in $d$-Au collisions measured by the PHENIX
Collaboration at the RHIC. In many cases, the fits with $a_0=1$
obtain several times larger $\chi^2$ than those with changeable
$a_0$.

The transverse momentum spectra, $d^{2}Y/dp_{\rm T}dy$, of
$J/\psi$ produced in Pb-Pb collisions at (a) 2.76 TeV and (b) 5.02
TeV in the rapidity interval $2.5<y<4$ are displayed in Figure 3,
where $Y$ denotes the yields. The symbols represent the
experimental data measured by the ALICE
Collaboration~\cite{22,23}. Panel (a) shows the spectra for three
centrality classes, 0--20$\%$, 20--40$\%$, and 40--90$\%$. Panel
(b) shows the spectra for seven centrality classes, 0--10\%,
10--20\%, 20--30\%, 30--40\%, 40--50\%, 50--60\%, and 60--90\%.
The sold (dot-dashed) curves are our fitting results by Eq. (6)
(with $a_0=1$). The values of fitting parameters are listed in
Table 3 with other information. One can see that Eq. (6) (with
changeable $a_0$) fits satisfactorily the experimental data in
Pb-Pb collisions measured by the ALICE Collaboration at the LHC.
In many cases, the fits with $a_0=1$ obtain several times larger
$\chi^2$ than those with changeable $a_0$.

The transverse momentum spectra, (a, b, i, and j)
$d^{2}\sigma/dp_{\rm T}dy$ or (c--h) $Bd^{2}\sigma/dp_{\rm T}dy$,
of (a, c, e, and g) prompt $J/\psi$, (b) $J/\psi$ from $b$, (d, f,
and h) nonprompt $J/\psi$, or (i and j) inclusive $J/\psi$
produced in $p$-Pb collisions at 5.02 TeV are given in Figure 4.
As can be seen in the figure, panels (a--h) show the spectra for
different rapidity intervals, while panels (i) and (j) show the
spectra for given rapidity interval and different centrality
classes. The symbols in panels (a and b), (c--f), (g and h), as
well as (i and j) represent the experimental data measured by the
LHCb~\cite{24}, CMS~\cite{25}, ATLAS~\cite{26}, and ALICE
Collaborations~\cite{27}, respectively. The solid (dot-dashed)
curves are our fitting results by Eq. (6) (with $a_0=1$). The
values of fitting parameters are listed in Table 4. It should be
noted here that, in Figures 4(d), 4(f), and 4(h), although
nonprompt $J/\psi$ is produced from the fragmentation of open
bottom hadron, it is also regarded as two contributors due to the
fact that open bottom hadron has two contributors. This is similar
to the view of point of string, in which two contributors form a
string. Then, the string is broken to produce a particle, and the
particle has two contributors. From Figure 4 one can see that Eq.
(6) (with changeable $a_0$) fits satisfactorily the experimental
data in $p$-Pb collisions measured by several collaborations at
the LHC. In many cases, the fits with $a_0=1$ obtain several times
larger $\chi^2$ than those with changeable $a_0$.

In Figure 5, we show the $J/\psi$ transverse momentum spectra,
(a--c and g--j) $Bd^{2}\sigma/(2\pi p_{\rm T}dp_{\rm T}dy)$, (d
and e) $Bd\sigma/dp_{\rm T}$, (f) $d\sigma/dp_{\rm T}$, and (k--p)
$d^{2}\sigma/dp_{\rm T}dy$ in $p$-$p(\overline{p})$ collisions at
center-of-mass energy $\sqrt{s}=$ (a) 200, (b) 500, and (c) 510
GeV, as well as (d) 1.8, (e) 1.96, (f) 2.76, (g) 5.02, and (h)
5.02 TeV, with different $y$ or $\eta$ (pseudorapidity) and other
selection conditions marked in the panels. The data symbols in
panels (a), (b and c), (d and e), (f), (g--j), and (k--p) are
quoted from the PHENIX~\cite{28,29,30}, STAR~\cite{31},
CDF~\cite{32,33}, LHCb~\cite{34} and ALICE~\cite{35},
CMS~\cite{25}, and LHCb Collaborations~\cite{36,37,38},
respectively. The solid (dot-dashed) curves are our fitting
results by Eq. (6) (with $a_0=1$). The values of fitting
parameters are listed in Table 5. One can see that Eq. (6) (with
changeable $a_0$) fits satisfactorily the experimental data in
$p$-$p$ collisions measured by several collaborations at the RHIC,
Tevatron, and LHC. In many cases, the fits with $a_0=1$ obtain
several times larger $\chi^2$ than those with changeable $a_0$.

\clearpage

\begin{figure*}
\begin{center}
\includegraphics[width=12.0cm]{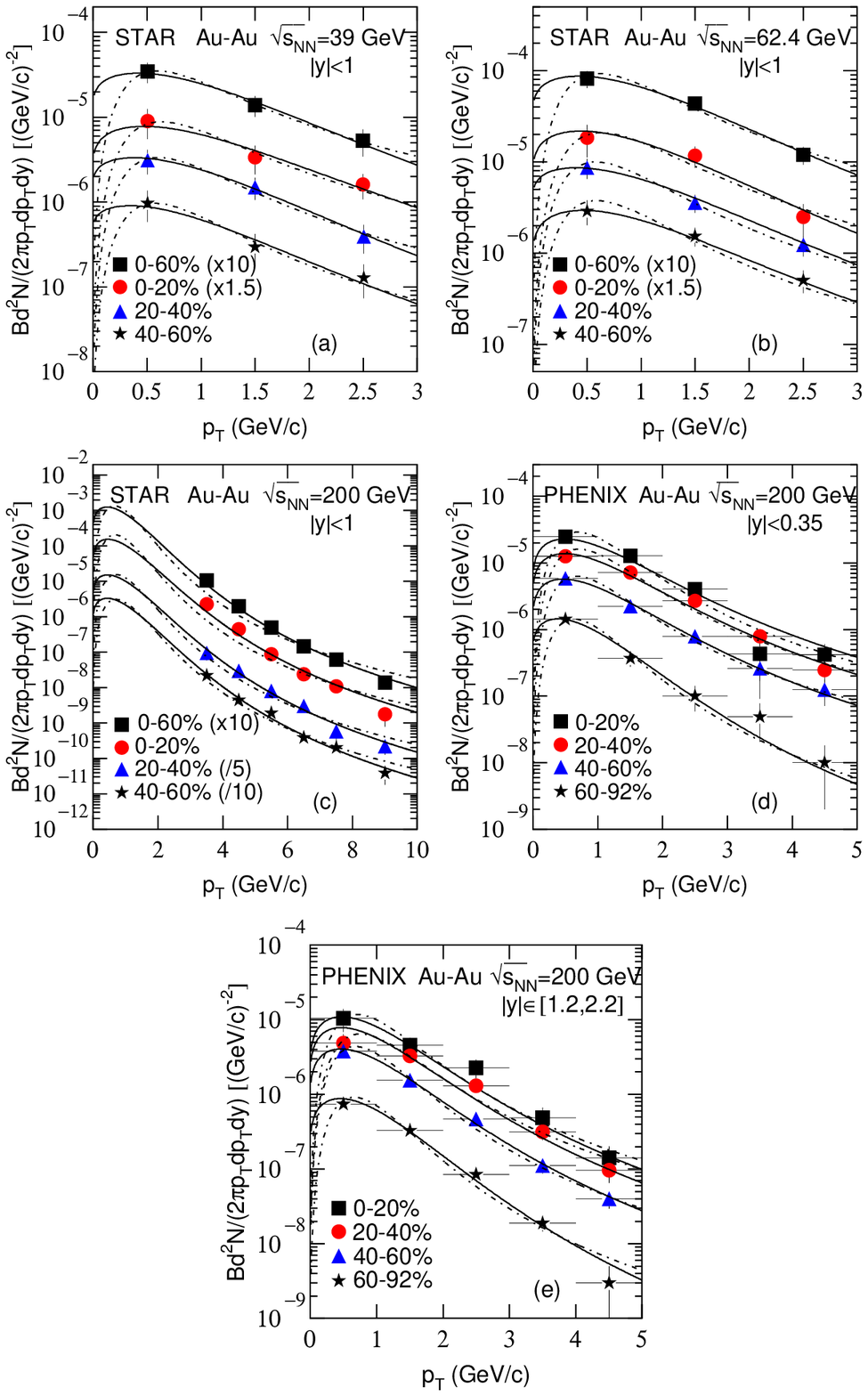}
\end{center}
\justifying\noindent {Figure 1. Transverse momentum spectra,
$Bd^{2}N/(2\pi p_{\rm T}dp_{\rm T}dy)$, of $J/\psi$ produced in
high energy Au-Au collisions with various centralities. Different
symbols in panels (a--c) show the spectra measured by the STAR
Collaboration~\cite{19} at $\sqrt{s_{\rm NN}}=$ (a) 39, (b) 62.4,
and (c) 200 GeV with $|y|<1$. Panels (d) and (e) are the spectra
measured by the PHENIX Collaboration~\cite{20} at 200 GeV with
$|y|<0.35$ and $y\in[1.2,2.2]$ respectively. The solid curves are
our fitting results by using Eq. (6) (with changeable $a_0$), and
the dot-dashed curves are our results refitted by Eq. (6) with
unchangeable $a_0=1$.}
\end{figure*}

\begin{widetext}
\begin{sidewaystable}
\vspace{9.0cm} \justifying\noindent {\scriptsize Table 1. Left
panel: Values of $n$, $T$, $a_{0}$, $N_{0}$, $\chi^2$, and ndof
corresponding to the solid curves in Figure 1 for Au-Au
collisions. Right panel: Values of $n$, $T$, $N_{0}$, $\chi^2$,
and ndof corresponding to the dot-dashed curves in Figure 1 for
Au-Au collisions, in which $a_0=1$.

\vspace{-0.35cm}

\begin{center}
\newcommand{\tabincell}[2]{\begin{tabular}{@{}#1@{}}#2\end{tabular}}
\begin{tabular} {ccccccccc|cccc}\\ \hline\hline
Figure   & Collab. &  $\sqrt{s_{\rm NN}}$ (GeV) & Selection & $n$ & $T$ (GeV) & $a_0$ & $N_0$ & $\chi^2$/ndof & $n$ & $T$ (GeV) & $N_0$ & $\chi^2$/ndof \\
\hline
Figure 1(a) & STAR    & $39   $           & 0--60\%  & $2.70\pm0.02$  & $0.278\pm0.001$ & $0.131\pm0.001$ & $(3.43\pm0.31)\times10^{-4}$ & $0.2/-$ & $1.56\pm0.01$ & $0.037\pm0.001$ & $(3.55\pm0.33)\times10^{-4}$ & $0.2/-$\\
          & Au-Au     & $|y|<1$           & 0--20\%  & $2.37\pm0.02$  & $0.297\pm0.001$ & $0.150\pm0.001$ & $(6.23\pm0.58)\times10^{-5}$ & $0.5/-$ & $1.62\pm0.01$ & $0.042\pm0.001$ & $(9.35\pm0.87)\times10^{-5}$ & $0.5/-$\\
          &           & $     $           & 20--40\% & $2.87\pm0.02$  & $0.278\pm0.001$ & $0.112\pm0.001$ & $(3.19\pm0.28)\times10^{-5}$ & $0.1/-$ & $1.65\pm0.01$ & $0.039\pm0.001$ & $(3.19\pm0.30)\times10^{-5}$ & $0.3/-$\\
          &           & $     $           & 40--60\% & $2.95\pm0.02$  & $0.252\pm0.001$ & $0.108\pm0.001$ & $(8.29\pm0.79)\times10^{-6}$ & $0.4/-$ & $1.67\pm0.01$ & $0.036\pm0.001$ & $(8.33\pm0.81)\times10^{-6}$ & $0.4/-$\\
Figure 1(b) & STAR    & $62.4 $           & 0--60\%  & $2.86\pm0.02$  & $0.285\pm0.001$ & $0.131\pm0.001$ & $(9.07\pm0.85)\times10^{-4}$ & $0.3/-$ & $1.60\pm0.01$ & $0.039\pm0.001$ & $(9.40\pm0.91)\times10^{-4}$ & $1/-$\\
          & Au-Au     & $|y|<1$           & 0--20\%  & $2.78\pm0.02$  & $0.275\pm0.001$ & $0.162\pm0.001$ & $(2.20\pm0.20)\times10^{-4}$ & $0.8/-$ & $1.66\pm0.01$ & $0.043\pm0.001$ & $(2.16\pm0.19)\times10^{-4}$ & $1/-$\\
          &           & $     $           & 20--40\% & $2.80\pm0.02$  & $0.271\pm0.001$ & $0.145\pm0.001$ & $(9.00\pm0.86)\times10^{-5}$ & $0.4/-$ & $1.69\pm0.01$ & $0.037\pm0.001$ & $(8.63\pm0.84)\times10^{-5}$ & $0.4/-$\\
          &           & $     $           & 40--60\% & $2.94\pm0.02$  & $0.266\pm0.001$ & $0.150\pm0.001$ & $(3.34\pm0.29)\times10^{-5}$ & $0.1/-$ & $1.70\pm0.01$ & $0.038\pm0.001$ & $(3.30\pm0.31)\times10^{-5}$ & $1/-$\\
Figure 1(c) & STAR    & $200  $           & 0--60\%  & $4.67\pm0.03$  & $0.169\pm0.001$ & $0.221\pm0.001$ & $(7.86\pm0.76)\times10^{-3}$ & $4/2$   & $3.94\pm0.03$ & $0.064\pm0.001$ & $(7.70\pm0.75)\times10^{-3}$ & $34/3$\\
          & Au-Au     & $|y|<1$           & 0--20\%  & $4.62\pm0.03$  & $0.183\pm0.001$ & $0.225\pm0.001$ & $(1.07\pm0.09)\times10^{-3}$ & $10/2$  & $3.99\pm0.03$ & $0.067\pm0.001$ & $(1.23\pm0.10)\times10^{-3}$ & $29/3$\\
          &           & $     $           & 20--40\% & $4.73\pm0.03$  & $0.180\pm0.001$ & $0.221\pm0.001$ & $(1.02\pm0.08)\times10^{-4}$ & $4/2$   & $4.08\pm0.04$ & $0.073\pm0.001$ & $(1.01\pm0.08)\times10^{-4}$ & $14/3$\\
          &           & $     $           & 40--60\% & $4.76\pm0.03$  & $0.177\pm0.001$ & $0.215\pm0.001$ & $(2.16\pm0.18)\times10^{-5}$ & $3/2$   & $4.12\pm0.04$ & $0.075\pm0.001$ & $(2.05\pm0.19)\times10^{-5}$ & $10/3$\\
Figure 1(d) & PHENIX  & $200     $        & 0--20\%  & $2.23\pm0.02$  & $0.227\pm0.001$ & $0.208\pm0.001$ & $(2.76\pm0.24)\times10^{-4}$ & $8/1$   & $2.28\pm0.02$ & $0.064\pm0.001$ & $(2.74\pm0.25)\times10^{-4}$ & $9/2$\\
          & Au-Au     & $|y|<0.35$        & 20--40\% & $2.28\pm0.02$  & $0.225\pm0.001$ & $0.204\pm0.001$ & $(1.61\pm0.14)\times10^{-4}$ & $3/1$   & $2.20\pm0.02$ & $0.062\pm0.001$ & $(1.71\pm0.15)\times10^{-4}$ & $4/2$\\
          &           & $        $        & 40--60\% & $2.35\pm0.02$  & $0.216\pm0.001$ & $0.195\pm0.001$ & $(6.17\pm0.59)\times10^{-5}$ & $0.7/1$ & $2.12\pm0.02$ & $0.055\pm0.001$ & $(6.24\pm0.60)\times10^{-5}$ & $1/2$\\
          &           & $        $        & 60--92\% & $2.95\pm0.02$  & $0.179\pm0.001$ & $0.170\pm0.001$ & $(1.04\pm0.10)\times10^{-5}$ & $1/1$   & $2.56\pm0.02$ & $0.053\pm0.001$ & $(1.01\pm0.09)\times10^{-5}$ & $1/2$\\
Figure 1(e) & PHENIX  & $200     $        & 0--20\%  & $2.51\pm0.02$  & $0.199\pm0.001$ & $0.217\pm0.001$ & $(1.60\pm0.13)\times10^{-5}$ & $3/1$   & $2.34\pm0.02$ & $0.064\pm0.001$ & $(1.53\pm0.13)\times10^{-5}$ & $3/2$\\
          & Au-Au     & $|y|\in[1.2,2.2]$ & 20--40\% & $2.54\pm0.02$  & $0.196\pm0.001$ & $0.215\pm0.001$ & $(7.51\pm0.71)\times10^{-5}$ & $8/1$   & $2.30\pm0.02$ & $0.071\pm0.001$ & $(7.46\pm0.73)\times10^{-5}$ & $8/2$\\
          &           & $               $ & 40--60\% & $2.68\pm0.02$  & $0.194\pm0.001$ & $0.213\pm0.001$ & $(3.70\pm0.34)\times10^{-5}$ & $1/1$   & $2.48\pm0.02$ & $0.060\pm0.001$ & $(3.72\pm0.35)\times10^{-5}$ & $2/2$\\
          &           & $               $ & 60--92\% & $3.23\pm0.02$  & $0.192\pm0.001$ & $0.211\pm0.001$ & $(7.16\pm0.69)\times10^{-6}$ & $3/1$   & $2.59\pm0.02$ & $0.058\pm0.001$ & $(7.03\pm0.68)\times10^{-6}$ & $6/2$\\
\hline
\end{tabular}%
\end{center}}
\end{sidewaystable}
\end{widetext}

\begin{figure*}[htbp]
\begin{center}
\includegraphics[width=12.0cm]{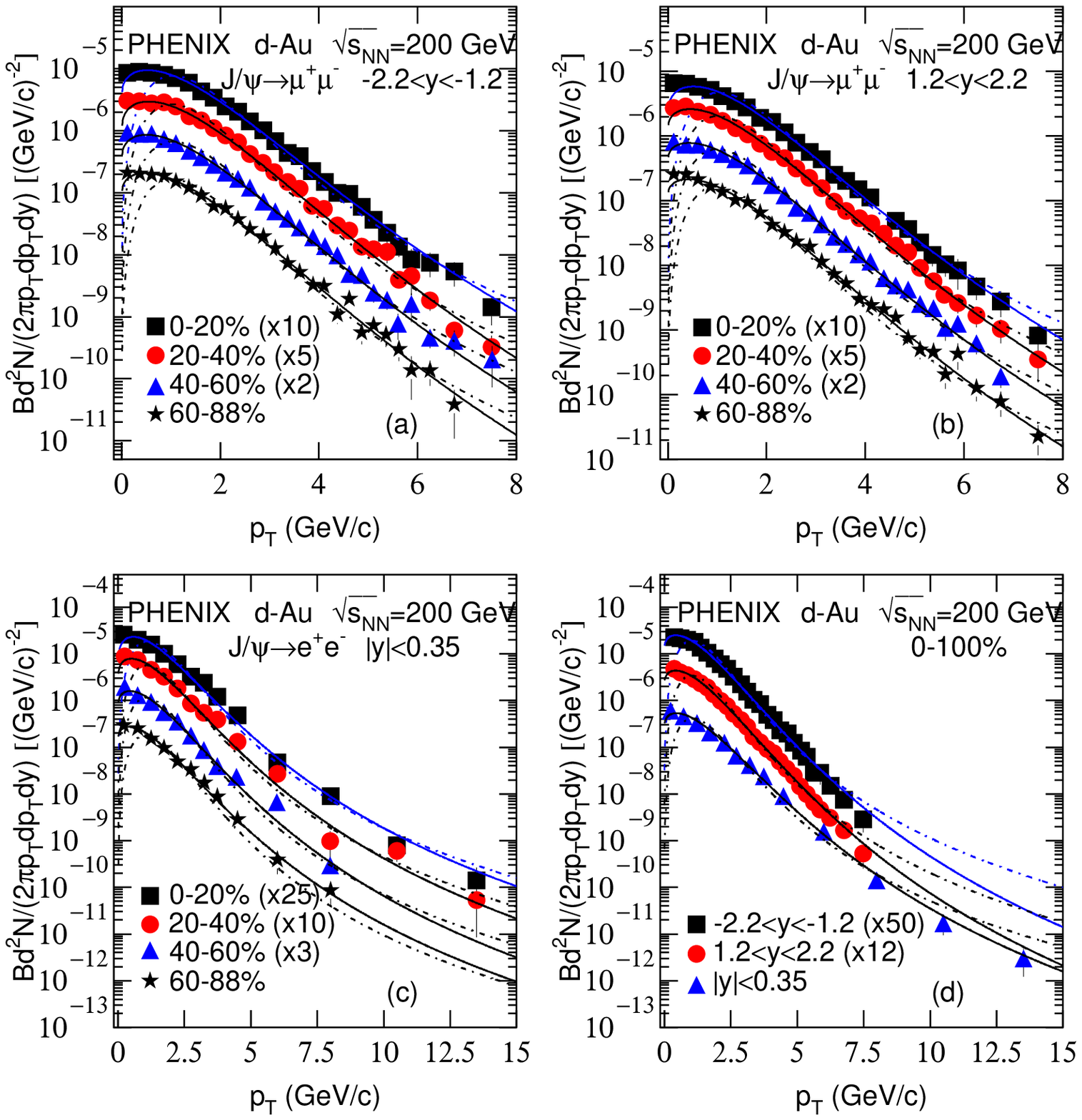}
\end{center}
\vspace{-0.25cm} \justifying\noindent {Figure 2. Transverse
momentum spectra, $Bd^{2}N/(2\pi p_{\rm T}dp_{\rm T}dy)$, of
$J/\psi$ produced in $d$-Au collisions at 200 GeV. Panels (a) and
(b) show the spectra of $J/\psi\rightarrow\mu^{+}\mu^{-}$ in the
backward and forward rapidity regions respectively, and panel (c)
shows the spectra of $J/\psi\rightarrow e^{+}e^{-}$ at
mid-rapidity $|y|<0.35$, with different centrality classes. Panel
(d) shows the spectra in minimum-bias collisions with different
rapidity intervals. The symbols represent the spectra measured by
the PHENIX Collaboration~\cite{21}, the solid curves are our
fitting results by Eq. (6), and the dot-dashed curves are our
results refitted by Eq. (6) with $a_0=1$.}
\end{figure*}

\begin{figure*}[htbp]
\begin{center}
\includegraphics[width=12.0cm]{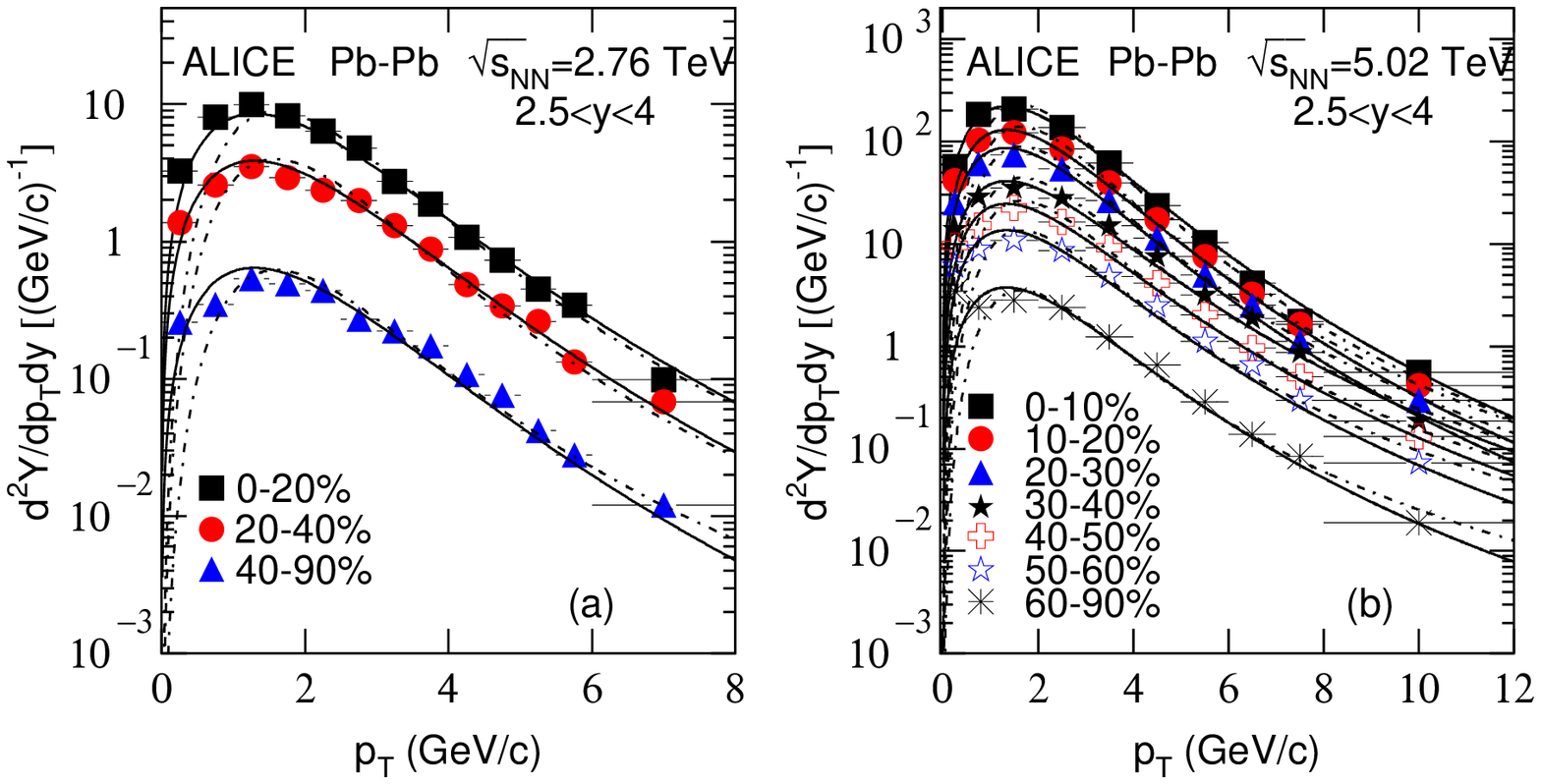}
\end{center}
\vspace{-0.25cm} \justifying\noindent {Figure 3. Transverse
momentum spectra, $d^{2}Y/dp_{\rm T}dy$, of $J/\psi$ produced in
Pb-Pb collisions at (a) 2.76 TeV and (b) 5.02 TeV, in the rapidity
interval $2.5<y<4$ and with different centrality classes. The
symbols represent the experimental data measured by the ALICE
Collaboration~\cite{22,23}, the solid curves are our fitting
results by Eq. (6), and the dot-dashed curves are our results
refitted by Eq. (6) with $a_0=1$.}
\end{figure*}

\clearpage

\begin{widetext}
\begin{sidewaystable}
\vspace{9.0cm}\justifying\noindent {\scriptsize Table 2. Left
panel: Values of $n$, $T$, $a_{0}$, $N_{0}$, $\chi^2$, and ndof
corresponding to the solid curves in Figure 2 for $d$-Au
collisions. Right panel: Values of $n$, $T$, $N_{0}$, $\chi^2$,
and ndof corresponding to the dot-dashed curves in Figure 2 for
$d$-Au collisions, in which $a_0=1$. \vspace{-.35cm}

\begin{center}
\newcommand{\tabincell}[2]{\begin{tabular}{@{}#1@{}}#2\end{tabular}}
\begin{tabular} {ccccccccc|cccc}\\ \hline\hline
Figure   & Collab. &  $\sqrt{s_{\rm NN}}$ (GeV) & Selection & $n$ & $T$ (GeV) & $a_0$ & $N_0$ & $\chi^2$/ndof & $n$ & $T$ (GeV) & $N_0$ & $\chi^2$/ndof \\
\hline
Figure 2(a) & PHENIX   & $200$         & 0--20\%       & $5.85\pm0.05$  & $0.310\pm0.001$ & $0.191\pm0.001$ & $(1.11\pm0.11)\times10^{-5}$ & $40/23$   & $5.35\pm0.05$ & $0.159\pm0.001$ & $(1.06\pm0.09)\times10^{-5}$ & $110/24$\\
          & $d$-Au     & $-2.2<y<-1.2$ & 20--40\%      & $5.88\pm0.05$  & $0.301\pm0.001$ & $0.165\pm0.001$ & $(6.90\pm0.64)\times10^{-6}$ & $119/23$  & $5.63\pm0.05$ & $0.154\pm0.001$ & $(6.89\pm0.67)\times10^{-6}$ & $137/24$\\
          &            & $    $        & 40--60\%      & $5.92\pm0.05$  & $0.299\pm0.001$ & $0.161\pm0.001$ & $(4.90\pm0.48)\times10^{-6}$ & $31/23$   & $5.61\pm0.05$ & $0.163\pm0.001$ & $(4.64\pm0.44)\times10^{-6}$ & $136/24$\\
          &            & $    $        & 60--88\%      & $5.99\pm0.05$  & $0.296\pm0.001$ & $0.160\pm0.001$ & $(2.27\pm0.21)\times10^{-6}$ & $36/22$   & $5.69\pm0.05$ & $0.157\pm0.001$ & $(2.16\pm0.20)\times10^{-6}$ & $128/23$\\
Figure 2(b) & PHENIX   & $200$         & 0--20\%       & $5.95\pm0.05$  & $0.311\pm0.001$ & $0.184\pm0.001$ & $(6.95\pm0.67)\times10^{-6}$ & $29/22$   & $5.20\pm0.05$ & $0.159\pm0.001$ & $(6.59\pm0.64)\times10^{-6}$ & $136/23$\\
          & $d$-Au     & $1.2<y<2.2$   & 20--40\%      & $6.08\pm0.05$  & $0.304\pm0.001$ & $0.175\pm0.001$ & $(5.50\pm0.50)\times10^{-6}$ & $42/23$   & $5.25\pm0.05$ & $0.153\pm0.001$ & $(5.41\pm0.52)\times10^{-6}$ & $194/23$\\
          &            & $    $        & 40--60\%      & $6.22\pm0.05$  & $0.299\pm0.001$ & $0.172\pm0.001$ & $(4.17\pm0.41)\times10^{-6}$ & $21/22$   & $5.29\pm0.05$ & $0.149\pm0.001$ & $(4.00\pm0.38)\times10^{-6}$ & $123/23$\\
          &            & $    $        & 60--88\%      & $6.36\pm0.06$  & $0.295\pm0.001$ & $0.168\pm0.001$ & $(2.18\pm0.20)\times10^{-6}$ & $39/23$   & $5.32\pm0.05$ & $0.145\pm0.001$ & $(2.05\pm0.19)\times10^{-6}$ & $187/24$\\
Figure 2(c) & PHENIX   & $200$         & 0--20\%       & $6.00\pm0.05$  & $0.352\pm0.001$ & $0.191\pm0.001$ & $(1.21\pm0.11)\times10^{-5}$ & $27/9$    & $5.31\pm0.05$ & $0.176\pm0.001$ & $(1.16\pm0.10)\times10^{-5}$ & $112/10$\\
          & $d$-Au     & $|y|<0.35$    & 20--40\%      & $6.05\pm0.05$  & $0.349\pm0.001$ & $0.178\pm0.001$ & $(9.39\pm0.91)\times10^{-6}$ & $26/9$    & $5.35\pm0.05$ & $0.171\pm0.001$ & $(8.54\pm0.83)\times10^{-6}$ & $116/10$\\
          &            & $    $        & 40--60\%      & $6.25\pm0.05$  & $0.338\pm0.001$ & $0.164\pm0.001$ & $(5.96\pm0.50)\times10^{-6}$ & $18/7$    & $5.40\pm0.05$ & $0.165\pm0.001$ & $(5.44\pm0.52)\times10^{-6}$ & $72/8$\\
          &            & $    $        & 60--88\%      & $6.29\pm0.05$  & $0.331\pm0.001$ & $0.157\pm0.001$ & $(2.93\pm0.27)\times10^{-6}$ & $10/7$    & $5.44\pm0.05$ & $0.161\pm0.001$ & $(2.55\pm0.24)\times10^{-6}$ & $70/8$\\
Figure 2(d) & PHENIX   & $200$         & $-2.2<y<-1.2$ & $5.89\pm0.05$  & $0.310\pm0.001$ & $0.148\pm0.001$ & $(5.37\pm0.48)\times10^{-6}$ & $50/23$   & $5.25\pm0.05$ & $0.157\pm0.001$ & $(5.25\pm0.50)\times10^{-6}$ & $206/24$\\
          & $d$-Au     & 0--100\%      & $1.2<y<2.2$   & $5.96\pm0.05$  & $0.315\pm0.001$ & $0.139\pm0.001$ & $(3.92\pm0.37)\times10^{-6}$ & $42/23$   & $5.25\pm0.05$ & $0.157\pm0.001$ & $(3.93\pm0.37)\times10^{-6}$ & $200/24$\\
          &            & $       $     & $|y|<0.35$    & $6.01\pm0.05$  & $0.362\pm0.001$ & $0.139\pm0.001$ & $(6.67\pm0.65)\times10^{-6}$ & $28/9$    & $4.89\pm0.04$ & $0.163\pm0.001$ & $(6.23\pm0.60)\times10^{-6}$ & $188/10$\\
\hline
\end{tabular}%
\end{center}}

\vspace{1.0cm} \justifying\noindent {\scriptsize Table 3. Left
panel: Values of $n$, $T$, $a_{0}$, $N_{0}$, $\chi^2$, and ndof
corresponding to the solid curves in Figure 3 for Pb-Pb
collisions. Right panel: Values of $n$, $T$, $N_{0}$, $\chi^2$,
and ndof corresponding to the dot-dashed curves in Figure 3 for
Pb-Pb collisions, in which $a_0=1$. \vspace{-0.35cm}

\begin{center}
\newcommand{\tabincell}[2]{\begin{tabular}{@{}#1@{}}#2\end{tabular}}
\begin{tabular} {ccccccccc|cccc}\\ \hline\hline
Figure & Collab. &  $\sqrt{s_{\rm NN}}$ (TeV) & Selection & $n$ & $T$ (GeV) & $a_0$ & $N_0$ & $\chi^2$/ndof & $n$ & $T$ (GeV) & $N_0$ & $\chi^2$/ndof \\
\hline
Figure 3(a) & ALICE & $2.76$    & 0--20\%  & $5.35\pm0.05$  & $0.409\pm0.001$ & $0.140\pm0.001$ & $(2.15\pm0.20)\times10^{1}$ & $14/9$  & $4.89\pm0.04$ & $0.202\pm0.001$ & $(2.10\pm0.18)\times10^{1}$ & $45/10$\\
          & Pb-Pb   & $2.5<y<4$ & 20--40\% & $5.37\pm0.05$  & $0.407\pm0.001$ & $0.137\pm0.001$ & $(9.70\pm0.95)\times10^{0}$ & $14/9$  & $4.91\pm0.04$ & $0.200\pm0.001$ & $(8.88\pm0.86)\times10^{0}$ & $70/10$\\
          &         & $   $     & 40--90\% & $5.39\pm0.05$  & $0.407\pm0.001$ & $0.135\pm0.001$ & $(1.61\pm0.14)\times10^{0}$ & $39/9$  & $4.93\pm0.04$ & $0.197\pm0.001$ & $(1.45\pm0.12)\times10^{0}$ & $64/10$\\
Figure 3(b) & ALICE & $5.02$    & 0--10\%  & $5.65\pm0.05$  & $0.431\pm0.001$ & $0.141\pm0.001$ & $(5.69\pm0.55)\times10^{2}$ & $47/6$  & $5.24\pm0.05$ & $0.215\pm0.001$ & $(5.18\pm0.50)\times10^{2}$ & $138/7$\\
          & Pb-Pb   & $2.5<y<4$ & 10--20\% & $5.68\pm0.05$  & $0.430\pm0.001$ & $0.141\pm0.001$ & $(3.34\pm0.32)\times10^{2}$ & $23/6$  & $4.91\pm0.04$ & $0.212\pm0.001$ & $(3.26\pm0.31)\times10^{2}$ & $108/7$\\
          &         & $    $    & 20--30\% & $5.70\pm0.05$  & $0.429\pm0.001$ & $0.140\pm0.001$ & $(2.23\pm0.20)\times10^{2}$ & $40/6$  & $4.76\pm0.04$ & $0.211\pm0.001$ & $(2.12\pm0.19)\times10^{2}$ & $107/7$\\
          &         & $    $    & 30--40\% & $4.56\pm0.04$  & $0.428\pm0.001$ & $0.139\pm0.001$ & $(1.12\pm0.10)\times10^{2}$ & $40/6$  & $4.42\pm0.04$ & $0.210\pm0.001$ & $(1.09\pm0.09)\times10^{2}$ & $119/7$\\
          &         & $    $    & 40--50\% & $4.57\pm0.04$  & $0.427\pm0.001$ & $0.136\pm0.001$ & $(6.67\pm0.65)\times10^{1}$ & $34/6$  & $4.34\pm0.04$ & $0.207\pm0.001$ & $(6.51\pm0.63)\times10^{1}$ & $86/7$\\
          &         & $    $    & 50--60\% & $4.58\pm0.04$  & $0.426\pm0.001$ & $0.135\pm0.001$ & $(3.67\pm0.35)\times10^{1}$ & $57/6$  & $4.17\pm0.04$ & $0.204\pm0.001$ & $(3.39\pm0.32)\times10^{1}$ & $136/7$\\
          &         & $    $    & 60--90\% & $4.60\pm0.04$  & $0.425\pm0.001$ & $0.135\pm0.001$ & $(1.01\pm0.08)\times10^{1}$ & $160/6$ & $4.13\pm0.06$ & $0.202\pm0.001$ & $(8.84\pm0.86)\times10^{0}$ & $267/7$\\
\hline
\end{tabular}%
\end{center}}
\end{sidewaystable}
\end{widetext}

\begin{figure*}[htbp]
\begin{center}
\includegraphics[width=8.5cm]{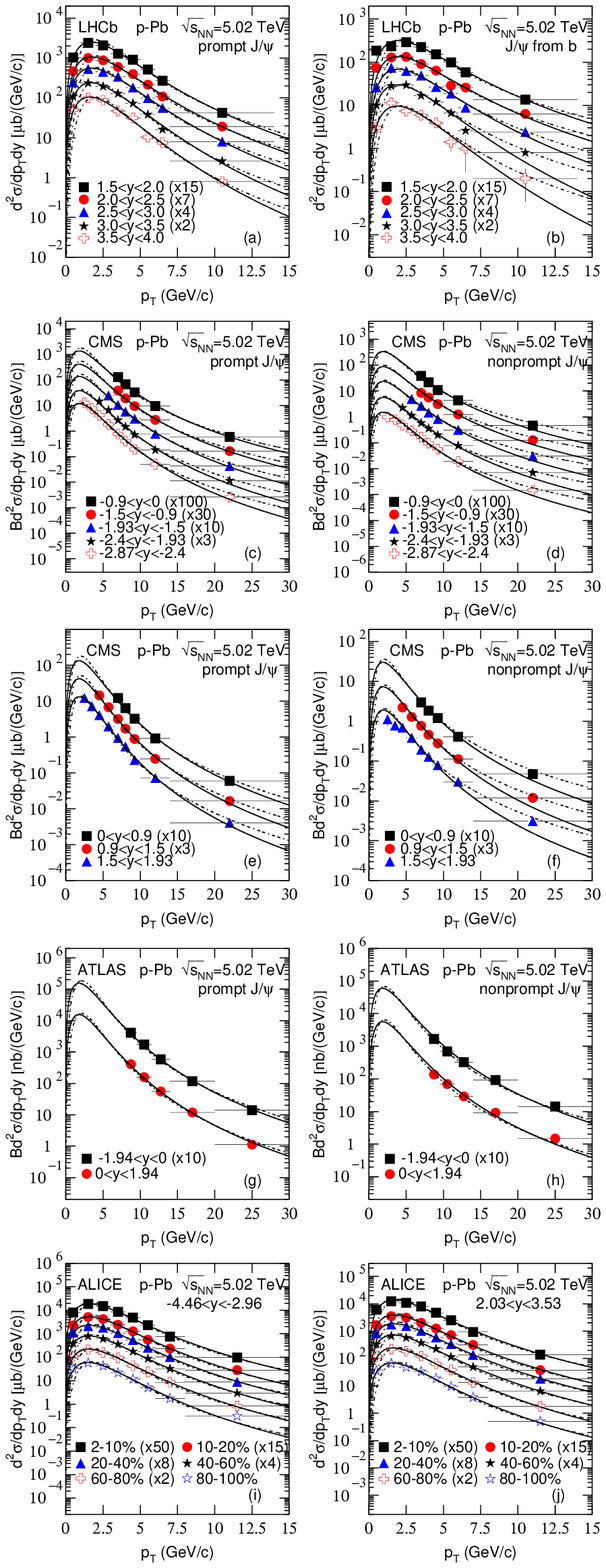}
\end{center}
\vspace{-0.25cm} \justifying\noindent {Figure 4. Transverse
momentum spectra, (a, b, i, and j) $d^{2}\sigma/dp_{\rm T}dy$ or
(c--h) $Bd^{2}\sigma/dp_{\rm T}dy$, of (a, c, e, and g) prompt
$J/\psi$, (b) $J/\psi$ from $b$, (d, f, and h) nonprompt $J/\psi$,
or (i and j) inclusive $J/\psi$ produced in $p$-Pb collisions at
5.02 TeV. Panels (a--h) show the spectra for different rapidity
intervals, while panels (i) and (j) show the spectra for given
rapidity interval and different centrality classes. The symbols in
panels (a and b), (c--f), (g and h), as well as (i and j)
represent the experimental data measured by the LHCb~\cite{24},
CMS~\cite{25}, ATLAS~\cite{26}, and ALICE~\cite{27}
Collaborations, respectively. The solid curves are our fitting
results by Eq. (6), and the dot-dashed curves are our results
refitted by Eq. (6) with $a_0=1$.}
\end{figure*}

\clearpage

\begin{widetext}
\begin{sidewaystable}
\vspace{9.0cm} \justifying\noindent {\scriptsize Table 4. Left
panel: Values of $n$, $T$, $a_{0}$, $\sigma_{0}$, $\chi^2$, and
ndof corresponding to the solid curves in Figure 4 for $p$-Pb
collisions. Right panel: Values of $n$, $T$, $\sigma_0$, $\chi^2$,
and ndof corresponding to the dot-dashed curves in Figure 4 for
$p$-Pb collisions, in which $a_0=1$. \vspace{-0.35cm}

\begin{center}
\newcommand{\tabincell}[2]{\begin{tabular}{@{}#1@{}}#2\end{tabular}}
\begin{tabular} {ccccccccc|cccc}\\ \hline\hline
Figure & Collab. &  $\sqrt{s_{\rm NN}}$ (TeV) & Selection & $n$ & $T$ (GeV) & $a_0$ & $\sigma_0$ ($\mu$b) & $\chi^2$/ndof & $n$ & $T$ (GeV) & $\sigma_0$ ($\mu$b) & $\chi^2$/ndof \\
\hline
Figure 4(a) & LHCb  & $5.02$ & $1.5<y<2.0$    & $4.55\pm0.03$  & $0.598\pm0.002$ & $0.172\pm0.001$ & $(5.97\pm0.57)\times10^{2}$  & $2/4$   & $4.27\pm0.04$ & $0.278\pm0.001$ & $(5.97\pm0.58)\times10^{2}$ & $41/5$\\
          & $p$-Pb  & $    $ & $2.0<y<2.5$    & $4.63\pm0.03$  & $0.612\pm0.002$ & $0.172\pm0.001$ & $(5.46\pm0.51)\times10^{2}$  & $7/4$   & $4.30\pm0.04$ & $0.284\pm0.001$ & $(5.15\pm0.50)\times10^{2}$ & $110/5$\\
          &         & $    $ & $2.5<y<3.0$    & $5.06\pm0.04$  & $0.622\pm0.002$ & $0.164\pm0.001$ & $(4.90\pm0.42)\times10^{2}$  & $5/4$   & $4.42\pm0.04$ & $0.287\pm0.001$ & $(4.57\pm0.44)\times10^{2}$ & $114/5$\\
          &         & $    $ & $3.0<y<3.5$    & $5.84\pm0.04$  & $0.622\pm0.002$ & $0.153\pm0.001$ & $(4.03\pm0.39)\times10^{2}$  & $9/4$   & $4.79\pm0.04$ & $0.289\pm0.001$ & $(3.74\pm0.36)\times10^{2}$ & $104/5$\\
          &         & $    $ & $3.5<y<4.0$    & $6.03\pm0.04$  & $0.624\pm0.002$ & $0.127\pm0.001$ & $(3.37\pm0.30)\times10^{2}$  & $10/4$  & $5.09\pm0.05$ & $0.292\pm0.001$ & $(3.21\pm0.30)\times10^{2}$ & $85/5$\\
Figure 4(b) & LHCb  & $5.02$ & $1.5<y<2.0$    & $4.01\pm0.03$  & $0.679\pm0.002$ & $0.226\pm0.001$ & $(9.00\pm0.88)\times10^{1}$  & $3/4$   & $3.64\pm0.03$ & $0.299\pm0.001$ & $(8.47\pm0.83)\times10^{1}$ & $9/5$\\
          & $p$-Pb  & $    $ & $2.0<y<2.5$    & $4.06\pm0.03$  & $0.680\pm0.002$ & $0.215\pm0.001$ & $(8.11\pm0.75)\times10^{1}$  & $11/4$  & $3.69\pm0.03$ & $0.302\pm0.001$ & $(7.74\pm0.75)\times10^{1}$ & $24/5$\\
          &         & $    $ & $2.5<y<3.0$    & $4.73\pm0.03$  & $0.688\pm0.002$ & $0.210\pm0.001$ & $(6.93\pm0.63)\times10^{1}$  & $4/4$   & $3.84\pm0.03$ & $0.306\pm0.001$ & $(6.58\pm0.64)\times10^{1}$ & $22/5$\\
          &         & $    $ & $3.0<y<3.5$    & $6.17\pm0.04$  & $0.694\pm0.002$ & $0.210\pm0.001$ & $(5.59\pm0.50)\times10^{1}$  & $6/4$   & $3.89\pm0.03$ & $0.309\pm0.001$ & $(5.27\pm0.51)\times10^{1}$ & $13/5$\\
          &         & $    $ & $3.5<y<4.0$    & $6.95\pm0.05$  & $0.697\pm0.002$ & $0.210\pm0.001$ & $(3.50\pm0.34)\times10^{1}$  & $4/4$   & $4.21\pm0.04$ & $0.312\pm0.001$ & $(3.41\pm0.33)\times10^{1}$ & $12/5$\\
Figure 4(c) & CMS   & $5.02$ & $-0.9<y<0$     & $4.96\pm0.04$  & $0.566\pm0.002$ & $0.254\pm0.001$ & $(4.74\pm0.45)\times10^{1}$  & $25/1$  & $4.86\pm0.04$ & $0.291\pm0.001$ & $(5.32\pm0.51)\times10^{1}$ & $59/2$\\
          & $p$-Pb  & $    $ & $-1.5<y<-0.9$  & $5.12\pm0.04$  & $0.576\pm0.002$ & $0.250\pm0.001$ & $(4.78\pm0.40)\times10^{1}$  & $18/1$  & $4.91\pm0.05$ & $0.295\pm0.001$ & $(5.42\pm0.52)\times10^{1}$ & $45/2$\\
          &         & $    $ & $-1.93<y<-1.5$ & $5.47\pm0.04$  & $0.591\pm0.002$ & $0.249\pm0.001$ & $(4.74\pm0.45)\times10^{1}$  & $23/2$  & $5.08\pm0.05$ & $0.306\pm0.001$ & $(5.05\pm0.48)\times10^{1}$ & $54/3$\\
          &         & $    $ & $-2.4<y<-1.93$ & $5.73\pm0.04$  & $0.593\pm0.002$ & $0.246\pm0.001$ & $(4.38\pm0.41)\times10^{1}$  & $62/3$  & $5.15\pm0.05$ & $0.314\pm0.001$ & $(4.33\pm0.41)\times10^{1}$ & $63/4$\\
          &         & $    $ & $-2.87<y<-2.4$ & $5.95\pm0.04$  & $0.598\pm0.002$ & $0.241\pm0.001$ & $(4.08\pm0.39)\times10^{1}$  & $39/5$  & $5.27\pm0.05$ & $0.317\pm0.001$ & $(3.99\pm0.37)\times10^{1}$ & $41/6$\\
Figure 4(d) & CMS   & $5.02$ & $-0.9<y<0$     & $4.49\pm0.03$  & $0.583\pm0.002$ & $0.278\pm0.001$ & $(1.24\pm0.10)\times10^{1}$  & $30/1$  & $4.06\pm0.04$ & $0.286\pm0.001$ & $(1.16\pm0.10)\times10^{1}$ & $31/2$\\
          & $p$-Pb  & $    $ & $-1.5<y<-0.9$  & $4.59\pm0.03$  & $0.604\pm0.002$ & $0.277\pm0.001$ & $(1.10\pm0.11)\times10^{1}$  & $23/1$  & $4.10\pm0.04$ & $0.291\pm0.001$ & $(1.11\pm0.08)\times10^{1}$ & $25/2$\\
          &         & $    $ & $-1.93<y<-1.5$ & $4.79\pm0.03$  & $0.623\pm0.002$ & $0.272\pm0.001$ & $(8.90\pm0.87)\times10^{0}$  & $23/2$  & $4.17\pm0.04$ & $0.297\pm0.001$ & $(8.82\pm0.86)\times10^{0}$ & $24/3$\\
          &         & $    $ & $-2.4<y<-1.93$ & $4.93\pm0.03$  & $0.627\pm0.002$ & $0.268\pm0.001$ & $(7.21\pm0.63)\times10^{0}$  & $24/3$  & $4.29\pm0.04$ & $0.302\pm0.001$ & $(6.92\pm0.67)\times10^{0}$ & $35/4$\\
          &         & $    $ & $-2.87<y<-2.4$ & $5.13\pm0.04$  & $0.630\pm0.002$ & $0.262\pm0.001$ & $(5.70\pm0.55)\times10^{0}$  & $18/5$  & $4.39\pm0.04$ & $0.313\pm0.001$ & $(4.96\pm0.48)\times10^{0}$ & $22/6$\\
Figure 4(e) & CMS   & $5.02$ & $0<y<0.9$      & $4.96\pm0.03$  & $0.562\pm0.002$ & $0.254\pm0.001$ & $(4.46\pm0.43)\times10^{1}$  & $23/1$  & $4.90\pm0.04$ & $0.294\pm0.001$ & $(5.49\pm0.53)\times10^{1}$ & $39/2$\\
          & $p$-Pb  & $    $ & $0.9<y<1.5$    & $5.19\pm0.04$  & $0.569\pm0.002$ & $0.251\pm0.001$ & $(4.84\pm0.40)\times10^{1}$  & $26/3$  & $4.95\pm0.05$ & $0.297\pm0.001$ & $(5.25\pm0.51)\times10^{1}$ & $26/4$\\
          &         & $    $ & $1.5<y<1.93$   & $5.43\pm0.04$  & $0.574\pm0.002$ & $0.245\pm0.001$ & $(4.51\pm0.43)\times10^{1}$  & $11/5$  & $5.02\pm0.05$ & $0.304\pm0.001$ & $(4.30\pm0.41)\times10^{1}$ & $23/6$\\
Figure 4(f) & CMS   & $5.02$ & $0<y<0.9$      & $4.43\pm0.03$  & $0.586\pm0.002$ & $0.281\pm0.001$ & $(1.13\pm0.10)\times10^{1}$  & $43/1$  & $4.08\pm0.04$ & $0.287\pm0.001$ & $(1.21\pm0.10)\times10^{1}$ & $47/2$\\
          & $p$-Pb  & $    $ & $0.9<y<1.5$    & $4.65\pm0.03$  & $0.606\pm0.002$ & $0.279\pm0.001$ & $(9.21\pm0.86)\times10^{0}$  & $42/3$  & $4.10\pm0.04$ & $0.292\pm0.001$ & $(9.20\pm0.90)\times10^{0}$ & $44/4$\\
          &         & $    $ & $1.5<y<1.93$   & $4.82\pm0.03$  & $0.619\pm0.002$ & $0.272\pm0.001$ & $(7.21\pm0.70)\times10^{0}$  & $32/5$  & $4.12\pm0.04$ & $0.305\pm0.001$ & $(7.23\pm0.70)\times10^{0}$ & $37/6$\\
Figure 4(g) & ATLAS & $5.02$ & $-1.94<y<0$    & $5.76\pm0.04$  & $0.583\pm0.002$ & $0.248\pm0.001$ & $(5.23\pm0.50)\times10^{1}$  & $2/1$   & $5.67\pm0.05$ & $0.329\pm0.001$ & $(5.63\pm0.54)\times10^{1}$ & $16/2$\\
          & $p$-Pb  & $    $ & $0<y<1.94$     & $5.85\pm0.04$  & $0.583\pm0.002$ & $0.241\pm0.001$ & $(5.23\pm0.50)\times10^{1}$  & $4/1$   & $5.79\pm0.05$ & $0.337\pm0.001$ & $(5.16\pm0.50)\times10^{1}$ & $21/2$\\
Figure 4(h) & ATLAS & $5.02$ & $-1.94<y<0$    & $5.35\pm0.04$  & $0.595\pm0.002$ & $0.259\pm0.001$ & $(2.11\pm0.19)\times10^{1}$  & $16/1$  & $5.22\pm0.05$ & $0.319\pm0.001$ & $(2.09\pm0.19)\times10^{1}$ & $18/2$\\
          & $p$-Pb  & $    $ & $0<y<1.94$     & $5.46\pm0.04$  & $0.607\pm0.002$ & $0.257\pm0.001$ & $(2.01\pm0.18)\times10^{1}$  & $28/1$  & $5.29\pm0.05$ & $0.324\pm0.001$ & $(1.99\pm0.18)\times10^{1}$ & $31/2$\\
Figure 4(i) & ALICE & $5.02$ & 2--10\%        & $5.08\pm0.04$  & $0.521\pm0.002$ & $0.184\pm0.001$ & $(1.18\pm0.56)\times10^{3}$  & $3/4$   & $4.87\pm0.04$ & $0.263\pm0.001$ & $(1.15\pm0.10)\times10^{3}$ & $60/5$\\
          & $p$-Pb  & $    $ & 10--20\%       & $5.13\pm0.04$  & $0.518\pm0.002$ & $0.184\pm0.001$ & $(1.09\pm0.10)\times10^{3}$  & $2/4$   & $4.90\pm0.04$ & $0.259\pm0.001$ & $(1.09\pm0.09)\times10^{3}$ & $68/5$\\
          &         & $    $ & 20--40\%       & $5.27\pm0.04$  & $0.517\pm0.002$ & $0.175\pm0.001$ & $(8.94\pm0.87)\times10^{2}$  & $4/4$   & $4.95\pm0.05$ & $0.260\pm0.001$ & $(8.58\pm0.84)\times10^{2}$ & $88/5$\\
          &         & $    $ & 40--60\%       & $5.36\pm0.04$  & $0.515\pm0.002$ & $0.175\pm0.001$ & $(6.11\pm0.55)\times10^{2}$  & $1/4$   & $4.99\pm0.05$ & $0.257\pm0.001$ & $(5.98\pm0.58)\times10^{2}$ & $84/5$\\
          &         & $    $ & 60--80\%       & $5.36\pm0.04$  & $0.515\pm0.002$ & $0.165\pm0.001$ & $(3.31\pm0.33)\times10^{2}$  & $2/4$   & $5.01\pm0.05$ & $0.257\pm0.001$ & $(3.12\pm0.29)\times10^{2}$ & $80/5$\\
          &         & $    $ & 80--100\%      & $5.56\pm0.04$  & $0.503\pm0.002$ & $0.153\pm0.001$ & $(1.74\pm0.17)\times10^{2}$  & $9/4$   & $5.03\pm0.05$ & $0.247\pm0.001$ & $(1.62\pm0.14)\times10^{2}$ & $70/5$\\
Figure 4(j) & ALICE & $5.02$ & 2--10\%        & $4.50\pm0.04$  & $0.561\pm0.002$ & $0.205\pm0.001$ & $(9.61\pm0.96)\times10^{2}$  & $10/4$  & $4.40\pm0.04$ & $0.280\pm0.001$ & $(9.06\pm0.89)\times10^{2}$ & $86/5$\\
          & $p$-Pb  & $    $ & 10--20\%       & $4.50\pm0.04$  & $0.560\pm0.002$ & $0.198\pm0.001$ & $(8.96\pm0.80)\times10^{2}$  & $6/4$   & $4.46\pm0.04$ & $0.278\pm0.001$ & $(8.61\pm0.84)\times10^{2}$ & $102/5$\\
          &         & $    $ & 20--40\%       & $4.51\pm0.04$  & $0.559\pm0.002$ & $0.192\pm0.001$ & $(8.12\pm0.78)\times10^{2}$  & $5/4$   & $4.49\pm0.04$ & $0.274\pm0.001$ & $(8.00\pm0.78)\times10^{2}$ & $126/5$\\
          &         & $    $ & 40--60\%       & $4.63\pm0.04$  & $0.558\pm0.002$ & $0.183\pm0.001$ & $(6.15\pm0.55)\times10^{2}$  & $4/4$   & $4.52\pm0.04$ & $0.271\pm0.001$ & $(5.91\pm0.57)\times10^{2}$ & $112/5$\\
          &         & $    $ & 60--80\%       & $4.84\pm0.04$  & $0.555\pm0.002$ & $0.171\pm0.001$ & $(3.96\pm0.39)\times10^{2}$  & $9/4$   & $4.55\pm0.04$ & $0.268\pm0.001$ & $(3.71\pm0.35)\times10^{2}$ & $111/5$\\
          &         & $    $ & 80--100\%      & $4.75\pm0.04$  & $0.548\pm0.002$ & $0.144\pm0.001$ & $(2.21\pm0.21)\times10^{2}$  & $8/4$   & $4.58\pm0.04$ & $0.264\pm0.001$ & $(2.06\pm0.19)\times10^{2}$ & $110/5$\\
\hline
\end{tabular}%
\end{center}}
\end{sidewaystable}
\end{widetext}

\begin{figure*}[htbp]
\begin{center}
\includegraphics[width=11.0cm]{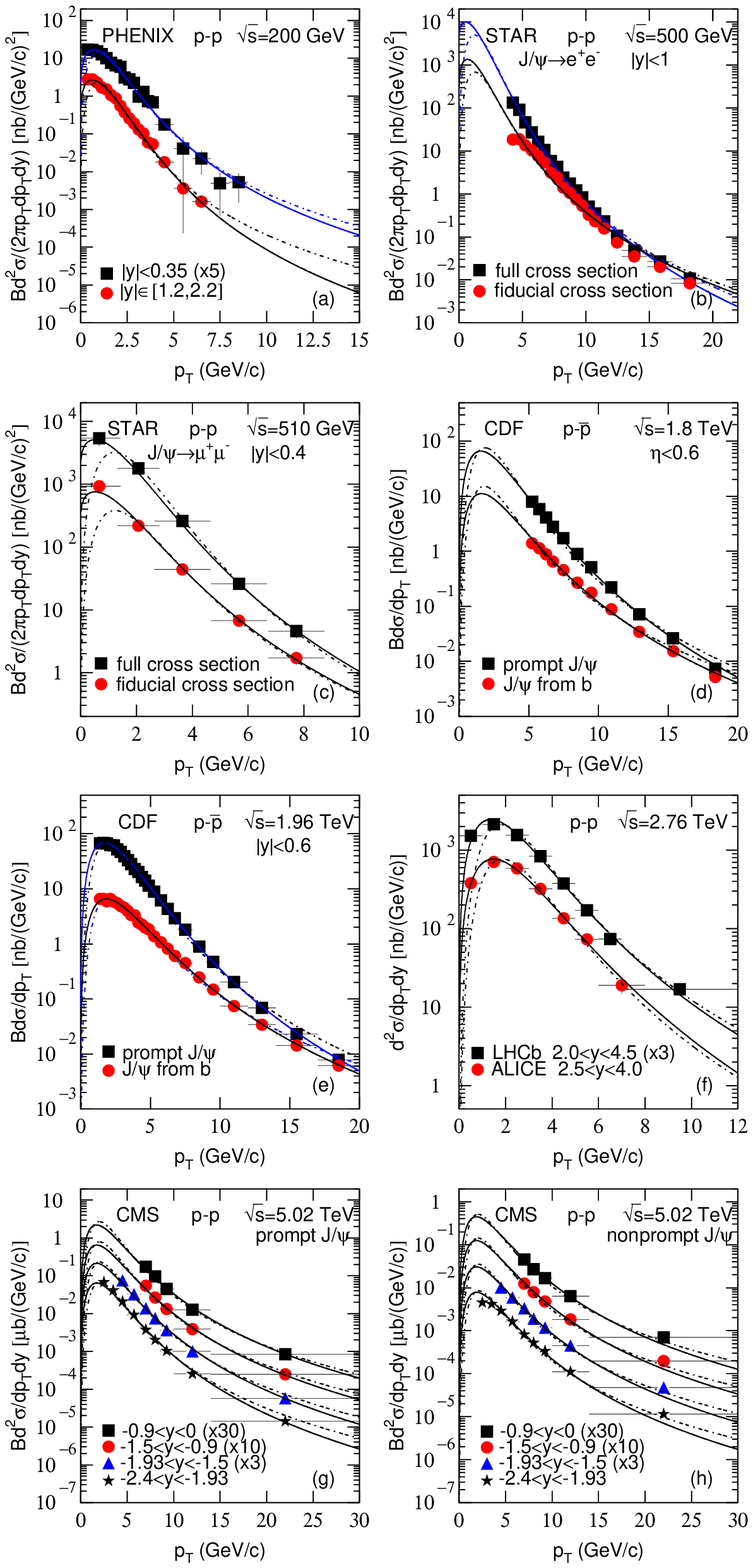}
\end{center}
\vspace{-0.25cm} \justifying\noindent {Figure 5. Transverse
momentum spectra, (a--c and g--j) $Bd^{2}\sigma/(2\pi p_{\rm
T}dp_{\rm T}dy)$, (d and e) $Bd\sigma/dp_{\rm T}$, (f)
$d\sigma/dp_{\rm T}$, and (k--p) $d^{2}\sigma/dp_{\rm T}dy$, of
$J/\psi$ produced in $p$-$p(\overline{p})$ collisions at different
energies. The data symbols in panels (a), (b and c), (d and e),
(f), (g--j), and (k--p) are quoted from the
PHENIX~\cite{28,29,30}, STAR~\cite{31}, CDF~\cite{32,33},
LHCb~\cite{34} and ALICE~\cite{35}, CMS~\cite{25}, and LHCb
Collaborations~\cite{36,37,38}, respectively. The solid curves are
our fitting results by Eq. (6), and the dot-dashed curves are our
results refitted by Eq. (6) with $a_0=1$.}
\end{figure*}

\begin{figure*}[htbp]
\begin{center}
\includegraphics[width=11.0cm]{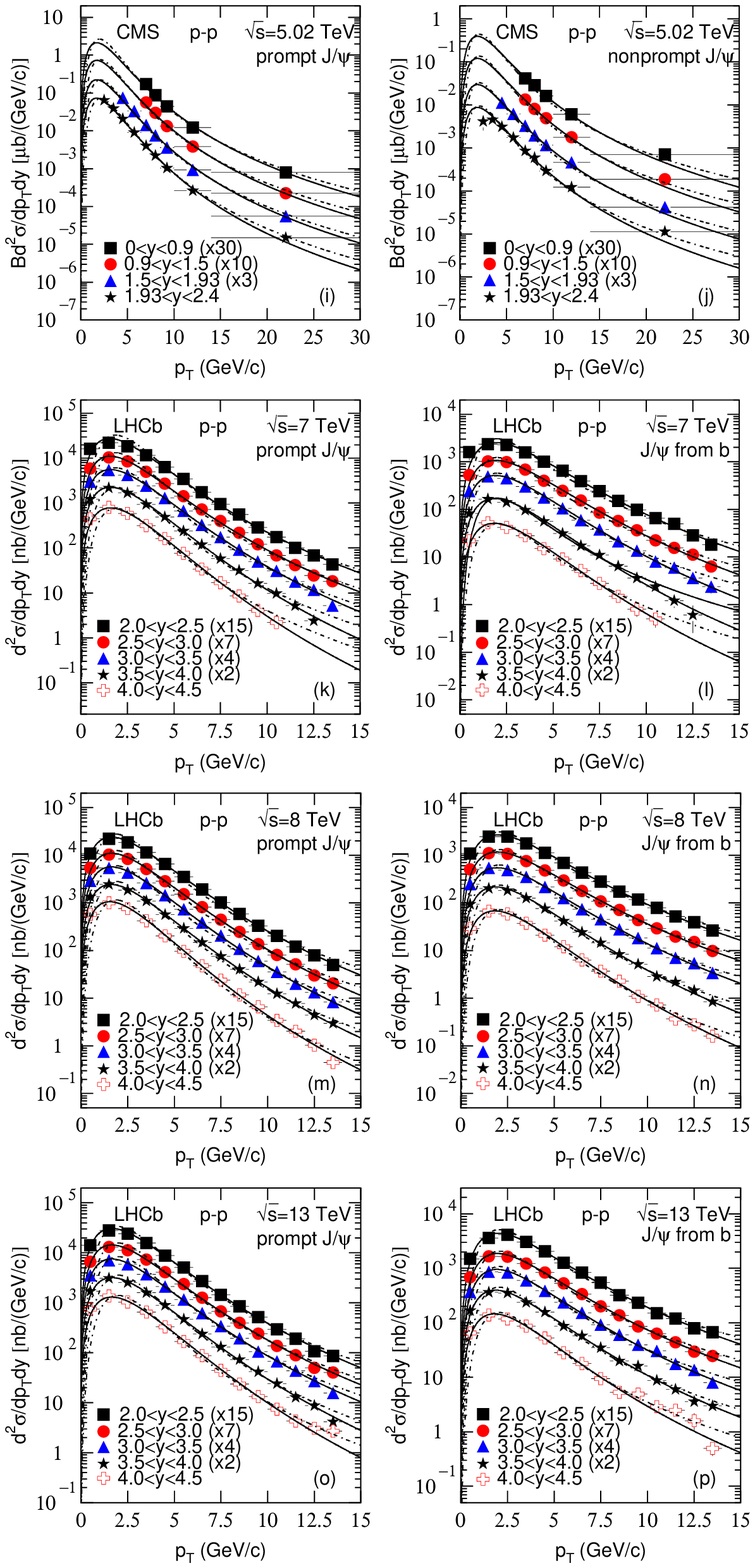}
\end{center}
\justifying\noindent {Figure 5. Continued. Panels (i--p) are
presented.}
\end{figure*}

\clearpage

\begin{widetext}
\begin{sidewaystable}
\vspace{9.0cm} \justifying\noindent {\scriptsize Table 5. Left
panel: Values of $n$, $T$, $a_{0}$, $\sigma_{0}$, $\chi^2$, and
ndof corresponding to the solid curves in Figure 5 for $p$-$p$
collisions. Right panel: Values of $n$, $T$, $\sigma_0$, $\chi^2$,
and ndof corresponding to the dot-dashed curves in Figure 5 for
$p$-$p$ collisions, in which $a_0=1$. \vspace{-.5cm}

\begin{center}
\newcommand{\tabincell}[2]{\begin{tabular}{@{}#1@{}}#2\end{tabular}}
\begin{tabular} {ccccccccc|cccc}\\ \hline\hline
Figure & Collab. &  $\sqrt{s_{\rm NN}}$ (TeV) & Selection & $n$ & $T$ (GeV) & $a_0$ & $\sigma_0$ ($\mu$b) & $\chi^2$/ndof & $n$ & $T$ (GeV) & $\sigma_0$ ($\mu$b) & $\chi^2$/ndof \\
\hline
Figure 5(a) & PHENIX & 0.2  & $|y|<0.35$             &  $4.42\pm0.03$  & $0.240\pm0.001$ & $0.363\pm0.001$ & $(4.15\pm0.38)\times10^{-2}$  & $19/17$  & $4.25\pm0.04$ & $0.139\pm0.001$ & $(4.15\pm0.40)\times10^{-2}$ & $47/18$\\
          & $p$-$p$  &      & $y\in[1.2,2.2]$        &  $5.26\pm0.04$  & $0.239\pm0.001$ & $0.306\pm0.001$ & $(2.76\pm0.26)\times10^{-2}$  & $26/15$  & $4.23\pm0.04$ & $0.119\pm0.001$ & $(2.65\pm0.25)\times10^{-2}$ & $108/16$\\
Figure 5(b) & STAR   & 0.5  & full cross section     &  $5.99\pm0.04$  & $0.350\pm0.001$ & $0.149\pm0.001$ & $(1.19\pm0.11)\times10^{2}$   & $11/15$  & $5.89\pm0.05$ & $0.211\pm0.001$ & $(9.36\pm0.91)\times10^{1}$  & $21/16$\\
          & $p$-$p$  &      & fiducial cross section &  $4.67\pm0.03$  & $0.389\pm0.001$ & $0.135\pm0.001$ & $(1.79\pm0.17)\times10^{1}$   & $143/15$ & $4.73\pm0.04$ & $0.212\pm0.001$ & $(1.48\pm0.13)\times10^{1}$  & $156/16$\\
Figure 5(c) & STAR   & 0.51 & full cross section     &  $4.60\pm0.03$  & $0.350\pm0.001$ & $0.149\pm0.001$ & $(6.58\pm0.65)\times10^{1}$   & $1/1$    & $5.20\pm0.05$ & $0.210\pm0.001$ & $(6.42\pm0.62)\times10^{1}$  & $6/2$\\
          & $p$-$p$  &      & fiducial cross section &  $3.73\pm0.02$  & $0.368\pm0.001$ & $0.146\pm0.001$ & $(1.11\pm0.11)\times10^{1}$   & $1/1$    & $4.14\pm0.04$ & $0.208\pm0.001$ & $(8.85\pm0.87)\times10^{0}$  & $10/2$\\
Figure 5(d) & CDF    & 1.8  & prompt $J/\psi$        &  $6.47\pm0.04$  & $0.535\pm0.002$ & $0.174\pm0.001$ & $(1.96\pm0.18)\times10^{-1}$  & $4/7$    & $5.77\pm0.05$ & $0.274\pm0.001$ & $(1.98\pm0.18)\times10^{-1}$ & $35/8$\\
          & $p$-$\overline{p}$ & & $J/\psi$ from $b$ &  $5.03\pm0.03$  & $0.529\pm0.002$ & $0.191\pm0.001$ & $(3.54\pm0.34)\times10^{-2}$  & $11/7$   & $4.68\pm0.04$ & $0.245\pm0.001$ & $(4.06\pm0.38)\times10^{-2}$ & $43/8$\\
Figure 5(e) & CDF    & 1.96 & prompt $J/\psi$        &  $5.13\pm0.04$  & $0.570\pm0.002$ & $0.285\pm0.001$ & $(2.30\pm0.23)\times10^{-2}$  & $4/22$   & $4.67\pm0.04$ & $0.299\pm0.001$ & $(2.23\pm0.20)\times10^{-2}$ & $38/23$\\
          & $p$-$\overline{p}$ & & $J/\psi$ from $b$ &  $6.67\pm0.04$  & $0.495\pm0.002$ & $0.322\pm0.001$ & $(2.05\pm0.20)\times10^{-1}$  & $6/22$   & $5.78\pm0.05$ & $0.287\pm0.001$ & $(1.87\pm0.17)\times10^{-1}$ & $57/23$\\
Figure 5(f) & LHCb   & 2.76 & $2.0<y<4.5$            &  $5.03\pm0.04$  & $0.452\pm0.002$ & $0.128\pm0.001$ & $(2.24\pm0.22)\times10^{0}$   & $6/4$    & $4.97\pm0.05$ & $0.252\pm0.001$ & $(1.97\pm0.16)\times10^{0}$  & $59/5$\\
          & ALICE    &      & $2.5<y<4.0$            &  $5.76\pm0.04$  & $0.505\pm0.002$ & $0.148\pm0.001$ & $(2.23\pm0.22)\times10^{0}$   & $3/3$    & $5.62\pm0.05$ & $0.272\pm0.001$ & $(2.03\pm0.18)\times10^{0}$  & $58/4$\\
Figure 5(g) & CMS    & 5.02 & $-0.9<y<0$             &  $4.85\pm0.03$  & $0.540\pm0.002$ & $0.244\pm0.001$ & $(2.47\pm0.24)\times10^{-1}$  & $30/1$   & $4.95\pm0.05$ & $0.292\pm0.001$ & $(2.69\pm0.25)\times10^{-1}$ & $39/2$\\
          & $p$-$p$  &      & $-1.5<y<-0.9$          &  $4.97\pm0.04$  & $0.553\pm0.002$ & $0.239\pm0.001$ & $(2.21\pm0.20)\times10^{-1}$  & $24/1$   & $4.99\pm0.05$ & $0.300\pm0.001$ & $(2.39\pm0.22)\times10^{-1}$ & $29/2$\\
          &          &      & $-1.93<y<-1.5$         &  $5.22\pm0.04$  & $0.557\pm0.002$ & $0.229\pm0.001$ & $(2.38\pm0.23)\times10^{-1}$  & $50/3$   & $5.13\pm0.05$ & $0.308\pm0.001$ & $(2.34\pm0.22)\times10^{-1}$ & $51/4$\\
          &          &      & $-2.4<y<-1.93$         &  $5.45\pm0.04$  & $0.560\pm0.002$ & $0.227\pm0.001$ & $(2.13\pm0.19)\times10^{-1}$  & $55/5$   & $5.18\pm0.05$ & $0.311\pm0.001$ & $(2.04\pm0.18)\times10^{-1}$ & $59/6$\\
Figure 5(h) & CMS    & 5.02 & $-0.9<y<0$             &  $4.39\pm0.03$  & $0.583\pm0.002$ & $0.267\pm0.001$ & $(5.60\pm0.50)\times10^{-2}$  & $67/1$   & $4.29\pm0.04$ & $0.289\pm0.001$ & $(5.56\pm0.54)\times10^{-2}$ & $74/2$\\
          & $p$-$p$  &      & $-1.5<y<-0.9$          &  $4.45\pm0.03$  & $0.591\pm0.002$ & $0.264\pm0.001$ & $(4.71\pm0.45)\times10^{-2}$  & $52/1$   & $4.34\pm0.04$ & $0.295\pm0.001$ & $(4.67\pm0.45)\times10^{-2}$ & $54/2$\\
          &          &      & $-1.93<y<-1.5$         &  $4.56\pm0.03$  & $0.601\pm0.002$ & $0.262\pm0.001$ & $(3.84\pm0.33)\times10^{-2}$  & $48/3$   & $4.39\pm0.04$ & $0.300\pm0.001$ & $(3.87\pm0.37)\times10^{-2}$ & $52/4$\\
          &          &      & $-2.4<y<-1.93$         &  $4.67\pm0.03$  & $0.616\pm0.002$ & $0.254\pm0.001$ & $(2.97\pm0.28)\times10^{-2}$  & $33/5$   & $4.44\pm0.04$ & $0.313\pm0.001$ & $(2.86\pm0.27)\times10^{-2}$ & $35/6$\\
Figure 5(i) & CMS    & 5.02 & $0<y<0.9$              &  $4.89\pm0.03$  & $0.542\pm0.002$ & $0.243\pm0.001$ & $(2.38\pm0.23)\times10^{-1}$  & $25/1$   & $4.94\pm0.05$ & $0.293\pm0.001$ & $(2.65\pm0.25)\times10^{-1}$ & $41/2$\\
          & $p$-$p$  &      & $0.9<y<1.5$            &  $5.07\pm0.04$  & $0.546\pm0.002$ & $0.239\pm0.001$ & $(2.42\pm0.22)\times10^{-1}$  & $27/1$   & $5.05\pm0.05$ & $0.307\pm0.001$ & $(2.33\pm0.21)\times10^{-1}$ & $54/2$\\
          &          &      & $1.5<y<1.93$           &  $5.30\pm0.04$  & $0.550\pm0.002$ & $0.239\pm0.001$ & $(2.46\pm0.24)\times10^{-1}$  & $62/3$   & $5.29\pm0.05$ & $0.320\pm0.001$ & $(2.28\pm0.21)\times10^{-1}$ & $48/4$\\
          &          &      & $1.93<y<2.4$           &  $5.65\pm0.04$  & $0.550\pm0.002$ & $0.238\pm0.001$ & $(2.37\pm0.21)\times10^{-1}$  & $18/5$   & $5.48\pm0.05$ & $0.326\pm0.001$ & $(2.19\pm0.20)\times10^{-1}$ & $18/6$\\
Figure 5(j) & CMS    & 5.02 & $0<y<0.9$              &  $4.38\pm0.03$  & $0.593\pm0.002$ & $0.269\pm0.001$ & $(4.99\pm0.40)\times10^{-2}$  & $78/1$   & $4.26\pm0.04$ & $0.294\pm0.001$ & $(4.88\pm0.48)\times10^{-2}$ & $82/2$\\
          & $p$-$p$  &      & $0.9<y<1.5$            &  $4.48\pm0.03$  & $0.597\pm0.002$ & $0.266\pm0.001$ & $(4.57\pm0.42)\times10^{-2}$  & $57/1$   & $4.34\pm0.04$ & $0.299\pm0.001$ & $(4.51\pm0.44)\times10^{-2}$ & $59/2$\\
          &          &      & $1.5<y<1.93$           &  $4.52\pm0.03$  & $0.599\pm0.002$ & $0.266\pm0.001$ & $(3.75\pm0.33)\times10^{-2}$  & $30/3$   & $4.38\pm0.04$ & $0.302\pm0.001$ & $(3.62\pm0.35)\times10^{-2}$ & $33/4$\\
          &          &      & $1.93<y<2.4$           &  $4.72\pm0.03$  & $0.603\pm0.002$ & $0.257\pm0.001$ & $(3.26\pm0.26)\times10^{-2}$  & $42/5$   & $4.44\pm0.04$ & $0.304\pm0.001$ & $(3.14\pm0.29)\times10^{-2}$ & $44/6$\\
Figure 5(k) & LHCb   & 7    & $2.0<y<2.5$            &  $5.46\pm0.04$  & $0.544\pm0.002$ & $0.149\pm0.001$ & $(5.69\pm0.56)\times10^{0}$   & $5/10$   & $5.07\pm0.05$ & $0.265\pm0.001$ & $(5.93\pm0.57)\times10^{0}$  & $30/11$\\
          & $p$-$p$  &      & $2.5<y<3.0$            &  $5.75\pm0.04$  & $0.563\pm0.002$ & $0.149\pm0.001$ & $(4.97\pm0.48)\times10^{0}$   & $3/10$   & $5.11\pm0.05$ & $0.265\pm0.001$ & $(5.14\pm0.49)\times10^{0}$  & $31/11$\\
          &          &      & $3.0<y<3.5$            &  $6.38\pm0.04$  & $0.578\pm0.002$ & $0.148\pm0.001$ & $(4.37\pm0.43)\times10^{0}$   & $4/10$   & $5.19\pm0.05$ & $0.267\pm0.001$ & $(4.14\pm0.40)\times10^{0}$  & $46/11$\\
          &          &      & $3.5<y<4.0$            &  $7.22\pm0.05$  & $0.583\pm0.002$ & $0.148\pm0.001$ & $(3.46\pm0.35)\times10^{0}$   & $4/9$    & $5.27\pm0.05$ & $0.269\pm0.001$ & $(3.05\pm0.28)\times10^{0}$  & $57/10$\\
          &          &      & $4.0<y<4.5$            &  $7.86\pm0.05$  & $0.585\pm0.002$ & $0.148\pm0.001$ & $(2.36\pm0.23)\times10^{0}$   & $2/7$    & $5.37\pm0.05$ & $0.271\pm0.001$ & $(2.07\pm0.19)\times10^{0}$  & $44/8$\\
Figure 5(l) & LHCb   & 7    & $2.0<y<2.5$            &  $5.06\pm0.03$  & $0.666\pm0.002$ & $0.226\pm0.001$ & $(6.62\pm0.66)\times10^{-1}$  & $3/10$   & $4.04\pm0.04$ & $0.274\pm0.001$ & $(6.59\pm0.63)\times10^{-1}$ & $13/11$\\
          & $p$-$p$  &      & $2.5<y<3.0$            &  $5.37\pm0.04$  & $0.668\pm0.002$ & $0.226\pm0.001$ & $(5.86\pm0.58)\times10^{-1}$  & $3/10$   & $4.10\pm0.04$ & $0.275\pm0.001$ & $(5.64\pm0.55)\times10^{-1}$ & $24/11$\\
          &          &      & $3.0<y<3.5$            &  $5.90\pm0.04$  & $0.668\pm0.002$ & $0.226\pm0.001$ & $(4.76\pm0.50)\times10^{-1}$  & $5/10$   & $4.19\pm0.04$ & $0.276\pm0.001$ & $(4.57\pm0.44)\times10^{-1}$ & $36/11$\\
          &          &      & $3.5<y<4.0$            &  $6.41\pm0.04$  & $0.677\pm0.002$ & $0.226\pm0.001$ & $(3.01\pm0.29)\times10^{-1}$  & $6/9$    & $4.28\pm0.04$ & $0.283\pm0.001$ & $(2.75\pm0.25)\times10^{-1}$ & $35/10$\\
          &          &      & $4.0<y<4.5$            &  $6.88\pm0.04$  & $0.681\pm0.002$ & $0.226\pm0.001$ & $(1.78\pm0.18)\times10^{-1}$  & $3/7$    & $4.39\pm0.04$ & $0.285\pm0.001$ & $(1.59\pm0.14)\times10^{-1}$ & $27/8$\\
\hline
\end{tabular}%
\end{center}}
\end{sidewaystable}
\end{widetext}

\clearpage

\begin{widetext}
\begin{sidewaystable}
\vspace{9.0cm} \justifying\noindent {\scriptsize Table 5.
Continued. The parameters for the curves in Figures 5(m), 5(n),
5(o), and 5(p) are listed. \vspace{-.5cm}

\begin{center}
\newcommand{\tabincell}[2]{\begin{tabular}{@{}#1@{}}#2\end{tabular}}
\begin{tabular} {ccccccccc|cccc}\\ \hline\hline
Figure & Collab. &  $\sqrt{s_{\rm NN}}$ (TeV) & Selection & $n$ & $T$ (GeV) & $a_0$ & $\sigma_0$ ($\mu$b) & $\chi^2$/ndof & $n$ & $T$ (GeV) & $\sigma_0$ ($\mu$b) & $\chi^2$/ndof \\
\hline
Figure 5(m) & LHCb   & 8    & $2.0<y<2.5$            &  $5.79\pm0.04$  & $0.602\pm0.002$ & $0.149\pm0.001$ & $(5.25\pm0.52)\times10^{0}$   & $4/10$  & $4.72\pm0.04$ & $0.260\pm0.001$ & $(5.15\pm0.50)\times10^{0}$  & $133/11$\\
          & $p$-$p$  &      & $2.5<y<3.0$            &  $6.11\pm0.04$  & $0.607\pm0.002$ & $0.146\pm0.001$ & $(5.11\pm0.48)\times10^{0}$   & $6/10$  & $4.78\pm0.04$ & $0.257\pm0.001$ & $(4.95\pm0.48)\times10^{0}$  & $153/11$\\
          &          &      & $3.0<y<3.5$            &  $6.51\pm0.04$  & $0.608\pm0.002$ & $0.141\pm0.001$ & $(4.56\pm0.42)\times10^{0}$   & $6/10$  & $4.97\pm0.05$ & $0.265\pm0.001$ & $(4.41\pm0.42)\times10^{0}$  & $168/11$\\
          &          &      & $3.5<y<4.0$            &  $6.86\pm0.04$  & $0.607\pm0.002$ & $0.133\pm0.001$ & $(3.84\pm0.38)\times10^{0}$   & $8/10$  & $5.31\pm0.05$ & $0.270\pm0.001$ & $(3.78\pm0.36)\times10^{0}$  & $157/11$\\
          &          &      & $4.0<y<4.5$            &  $7.98\pm0.05$  & $0.613\pm0.002$ & $0.133\pm0.001$ & $(3.15\pm0.30)\times10^{0}$   & $17/10$ & $5.91\pm0.06$ & $0.276\pm0.001$ & $(3.01\pm0.28)\times10^{0}$  & $181/11$\\
Figure 5(n) & LHCb   & 8    & $2.0<y<2.5$            &  $4.88\pm0.03$  & $0.660\pm0.002$ & $0.247\pm0.001$ & $(7.10\pm0.66)\times10^{-1}$  & $4/10$  & $4.04\pm0.04$ & $0.296\pm0.001$ & $(7.05\pm0.69)\times10^{-1}$ & $42/11$\\
          & $p$-$p$  &      & $2.5<y<3.0$            &  $5.03\pm0.04$  & $0.673\pm0.002$ & $0.242\pm0.001$ & $(6.55\pm0.65)\times10^{-1}$  & $7/10$  & $4.16\pm0.04$ & $0.298\pm0.001$ & $(6.34\pm0.62)\times10^{-1}$ & $62/11$\\
          &          &      & $3.0<y<3.5$            &  $5.48\pm0.04$  & $0.673\pm0.002$ & $0.242\pm0.001$ & $(5.20\pm0.47)\times10^{-1}$  & $8/10$  & $4.39\pm0.04$ & $0.300\pm0.001$ & $(5.11\pm0.49)\times10^{-1}$ & $58/11$\\
          &          &      & $3.5<y<4.0$            &  $5.99\pm0.04$  & $0.681\pm0.002$ & $0.214\pm0.001$ & $(3.92\pm0.39)\times10^{-1}$  & $12/10$ & $4.67\pm0.04$ & $0.302\pm0.001$ & $(3.73\pm0.35)\times10^{-1}$ & $120/11$\\
          &          &      & $4.0<y<4.5$            &  $6.69\pm0.04$  & $0.682\pm0.002$ & $0.214\pm0.001$ & $(2.37\pm0.23)\times10^{-1}$  & $8/10$  & $4.98\pm0.05$ & $0.304\pm0.001$ & $(2.22\pm0.20)\times10^{-1}$ & $94/11$\\
Figure 5(o) & LHCb   & 13   & $2.0<y<2.5$            &  $5.57\pm0.04$  & $0.615\pm0.002$ & $0.145\pm0.001$ & $(6.91\pm0.66)\times10^{0}$   & $21/10$ & $4.87\pm0.05$ & $0.288\pm0.001$ & $(6.70\pm0.65)\times10^{0}$  & $362/11$\\
          & $p$-$p$  &      & $2.5<y<3.0$            &  $5.61\pm0.04$  & $0.619\pm0.002$ & $0.138\pm0.001$ & $(6.91\pm0.68)\times10^{0}$   & $19/10$ & $4.92\pm0.05$ & $0.292\pm0.001$ & $(6.72\pm0.65)\times10^{0}$  & $360/11$\\
          &          &      & $3.0<y<3.5$            &  $5.81\pm0.04$  & $0.623\pm0.002$ & $0.130\pm0.001$ & $(6.30\pm0.60)\times10^{0}$   & $12/10$ & $5.03\pm0.05$ & $0.294\pm0.001$ & $(6.24\pm0.60)\times10^{0}$  & $320/11$\\
          &          &      & $3.5<y<4.0$            &  $6.89\pm0.04$  & $0.658\pm0.002$ & $0.116\pm0.001$ & $(5.31\pm0.50)\times10^{0}$   & $16/10$ & $5.31\pm0.05$ & $0.296\pm0.001$ & $(5.21\pm0.50)\times10^{0}$  & $243/11$\\
          &          &      & $4.0<y<4.5$            &  $7.61\pm0.05$  & $0.661\pm0.002$ & $0.116\pm0.001$ & $(4.27\pm0.41)\times10^{0}$   & $19/10$ & $5.45\pm0.05$ & $0.297\pm0.001$ & $(4.04\pm0.38)\times10^{0}$  & $153/11$\\
Figure 5(p) & LHCb   & 13   & $2.0<y<2.5$            &  $4.30\pm0.03$  & $0.668\pm0.002$ & $0.247\pm0.001$ & $(1.22\pm0.11)\times10^{0}$   & $5/10$  & $3.73\pm0.03$ & $0.282\pm0.001$ & $(1.22\pm0.10)\times10^{0}$  & $46/11$\\
          & $p$-$p$  &      & $2.5<y<3.0$            &  $4.44\pm0.03$  & $0.669\pm0.002$ & $0.239\pm0.001$ & $(1.09\pm0.11)\times10^{0}$   & $9/10$  & $3.78\pm0.03$ & $0.287\pm0.001$ & $(1.05\pm0.09)\times10^{0}$  & $134/11$\\
          &          &      & $3.0<y<3.5$            &  $4.72\pm0.03$  & $0.665\pm0.002$ & $0.229\pm0.001$ & $(9.41\pm0.90)\times10^{-1}$  & $11/10$ & $3.97\pm0.04$ & $0.289\pm0.001$ & $(9.27\pm0.91)\times10^{-1}$ & $146/11$\\
          &          &      & $3.5<y<4.0$            &  $4.92\pm0.04$  & $0.669\pm0.002$ & $0.192\pm0.001$ & $(7.60\pm0.73)\times10^{-1}$  & $10/10$ & $4.17\pm0.04$ & $0.291\pm0.001$ & $(7.50\pm0.73)\times10^{-1}$ & $117/11$\\
          &          &      & $4.0<y<4.5$            &  $5.51\pm0.04$  & $0.673\pm0.002$ & $0.169\pm0.001$ & $(5.35\pm0.52)\times10^{-1}$  & $13/10$ & $4.25\pm0.04$ & $0.292\pm0.001$ & $(5.09\pm0.49)\times10^{-1}$ & $82/11$\\
\hline
\end{tabular}%
\end{center}}
\end{sidewaystable}
\end{widetext}

\clearpage

\begin{figure*}[htbp]
\begin{center}
\includegraphics[width=12.0cm]{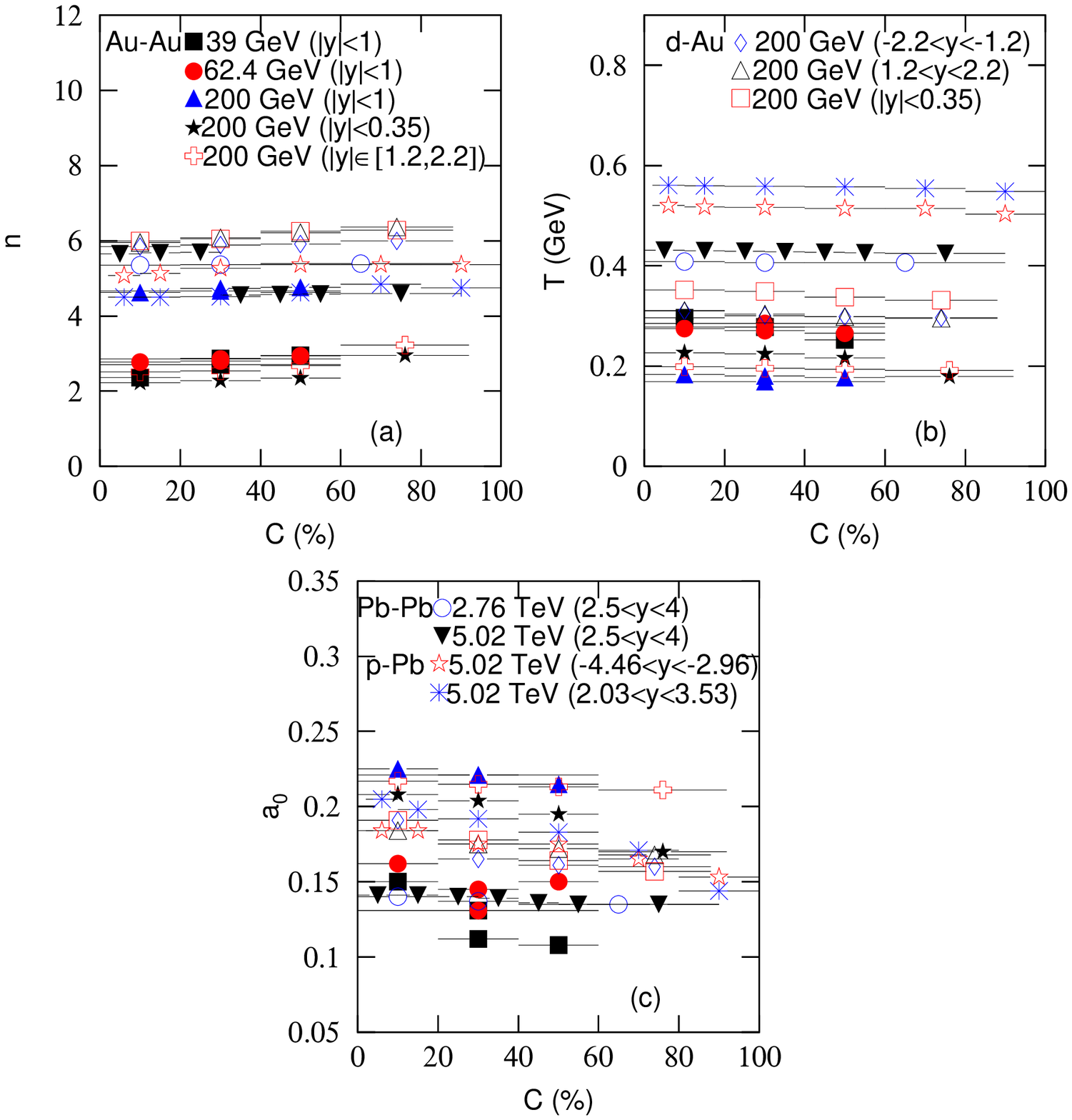}
\end{center}
\justifying\noindent {Figure 6. Dependences of power exponent $n$
(a), effective temperature $T$ (b), revised index $a_{0}$ (c) on
centrality percentage $C$ in Au-Au, $d$-Au, Pb-Pb, and $p$-Pb
collisions at the RHIC and LHC. The symbols represent the
parameter values listed in Tables 1--4 and extracted from Figures
1--4.}
\end{figure*}

\begin{figure*}[htbp]
\begin{center}
\includegraphics[width=12.0cm]{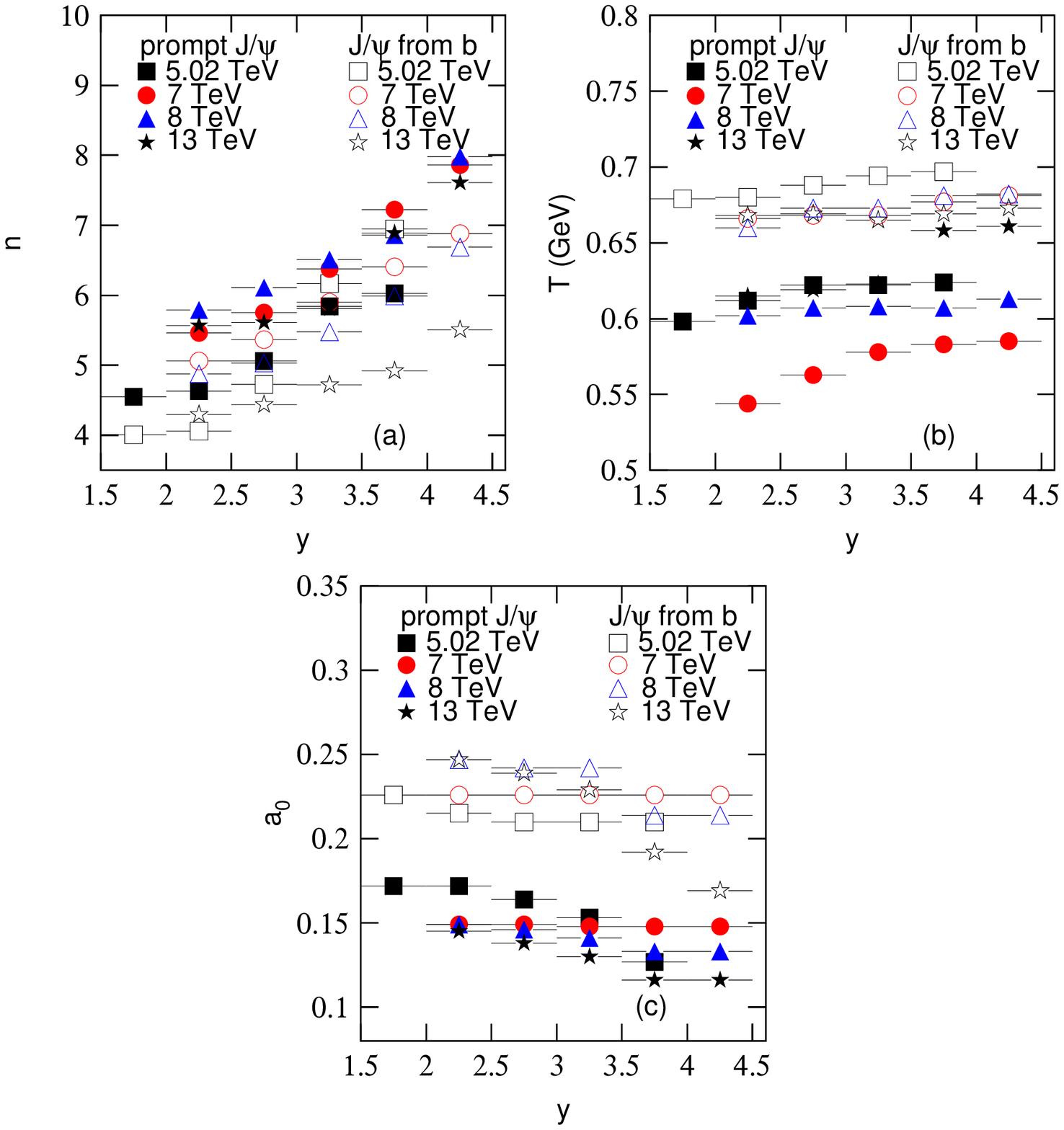}
\end{center}
\justifying\noindent {Figure 7. Dependences of (a) $n$, (b) $T$,
and (c) $a_{0}$ on rapidity $y$ in the forward region, which are
extracted from the spectra of prompt $J/\psi$ and $J/\psi$ from
$b$ in $p$-$p$ collisions at the LHC (5.02, 7, 8, and 13 TeV). The
symbols represent the parameter values listed in Table 5 and
obtained from Figure 5.}
\end{figure*}

\begin{figure*}[htbp]
\begin{center}
\includegraphics[width=12.0cm]{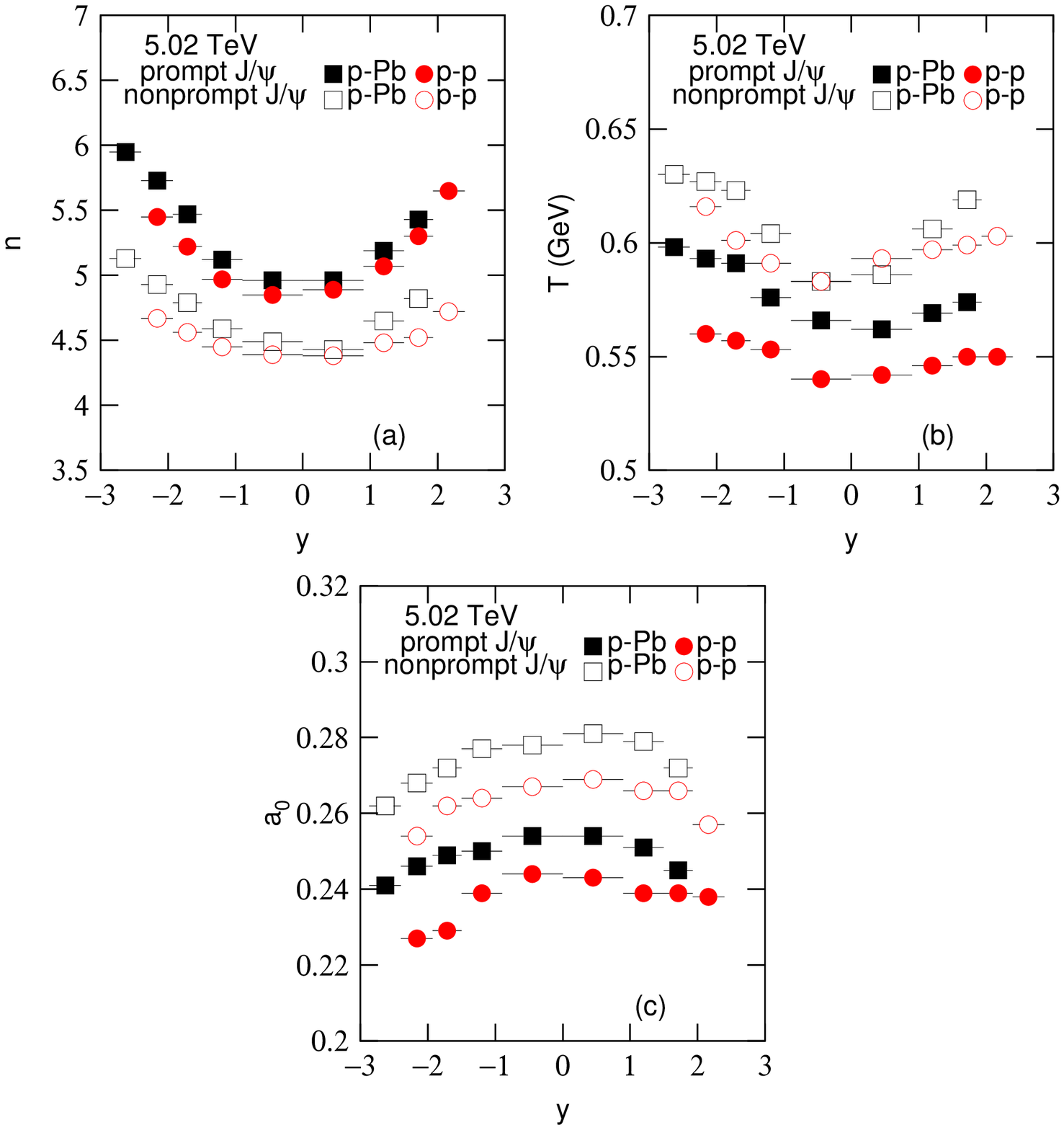}
\end{center}
\justifying\noindent {Figure 8. Dependences of (a) $n$, (b) $T$,
and (c) $a_{0}$ on $y$ in the forward and backward regions, which
are extracted from the spectra of prompt $J/\psi$ and nonprompt
$J/\psi$ in $p$-Pb and $p$-$p$ collisions at 5.02 TeV. The symbols
represent the parameter values listed in Tables 4 and 5 and
obtained from Figures 4 and 5.}
\end{figure*}

\begin{figure*}[htbp]
\begin{center}
\includegraphics[width=12.0cm]{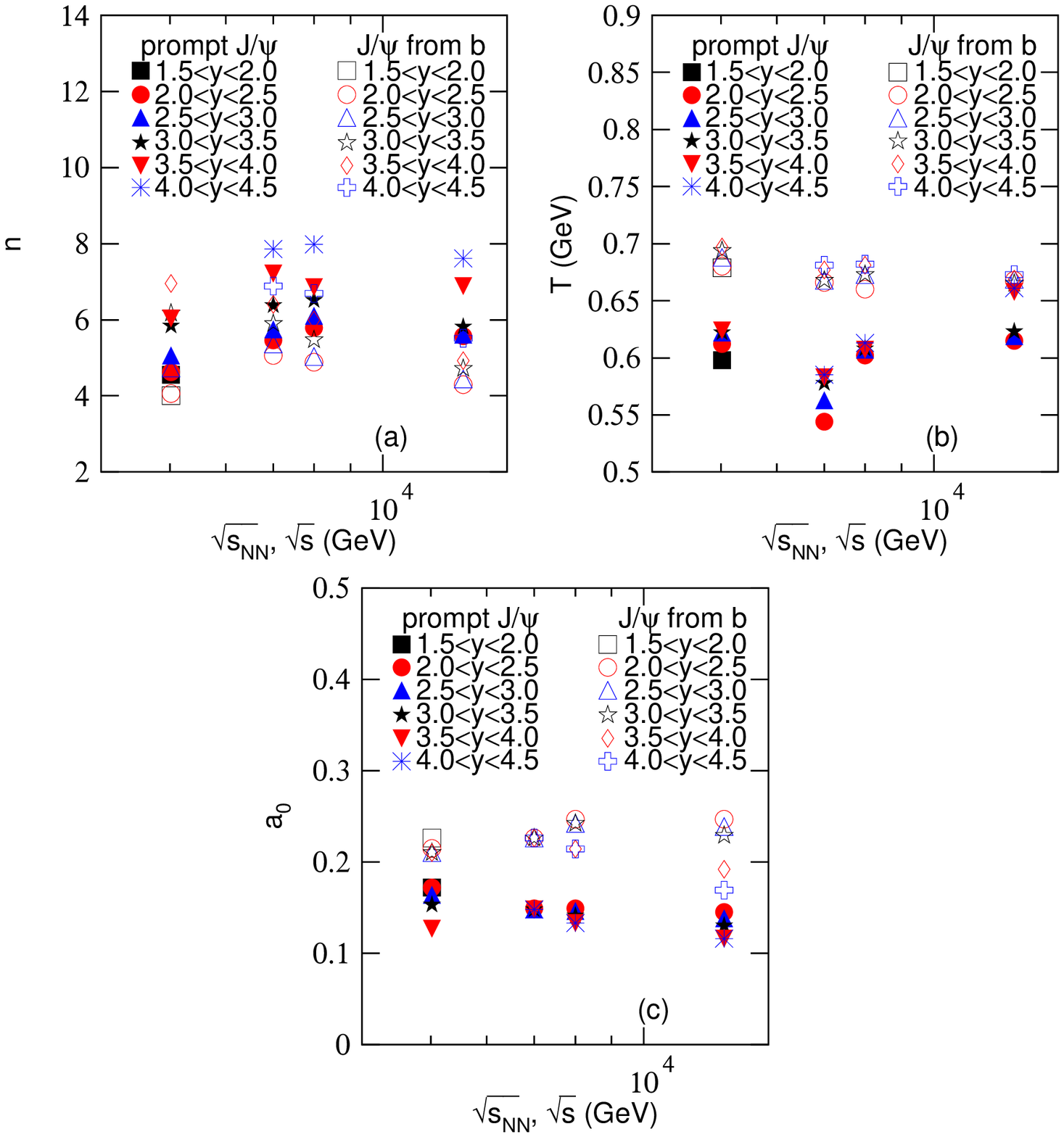}
\end{center}
\justifying\noindent {Figure 9. Dependences of (a) $n$, (b) $T$,
and (c) $a_{0}$ on energy $\sqrt{s_{\rm NN}}$ or $\sqrt{s}$ in the
forward rapidity region, which are extracted from the spectra of
prompt $J/\psi$ and $J/\psi$ from $b$ in $p$-Pb and $p$-$p$
collisions at the LHC (5.02, 7, 8, and 13 TeV). The symbols
represent the parameter values listed in Tables 4 and 5 and
obtained from Figures 4 and 5. }
\end{figure*}

\begin{figure*}[htbp]
\begin{center}
\includegraphics[width=12.0cm]{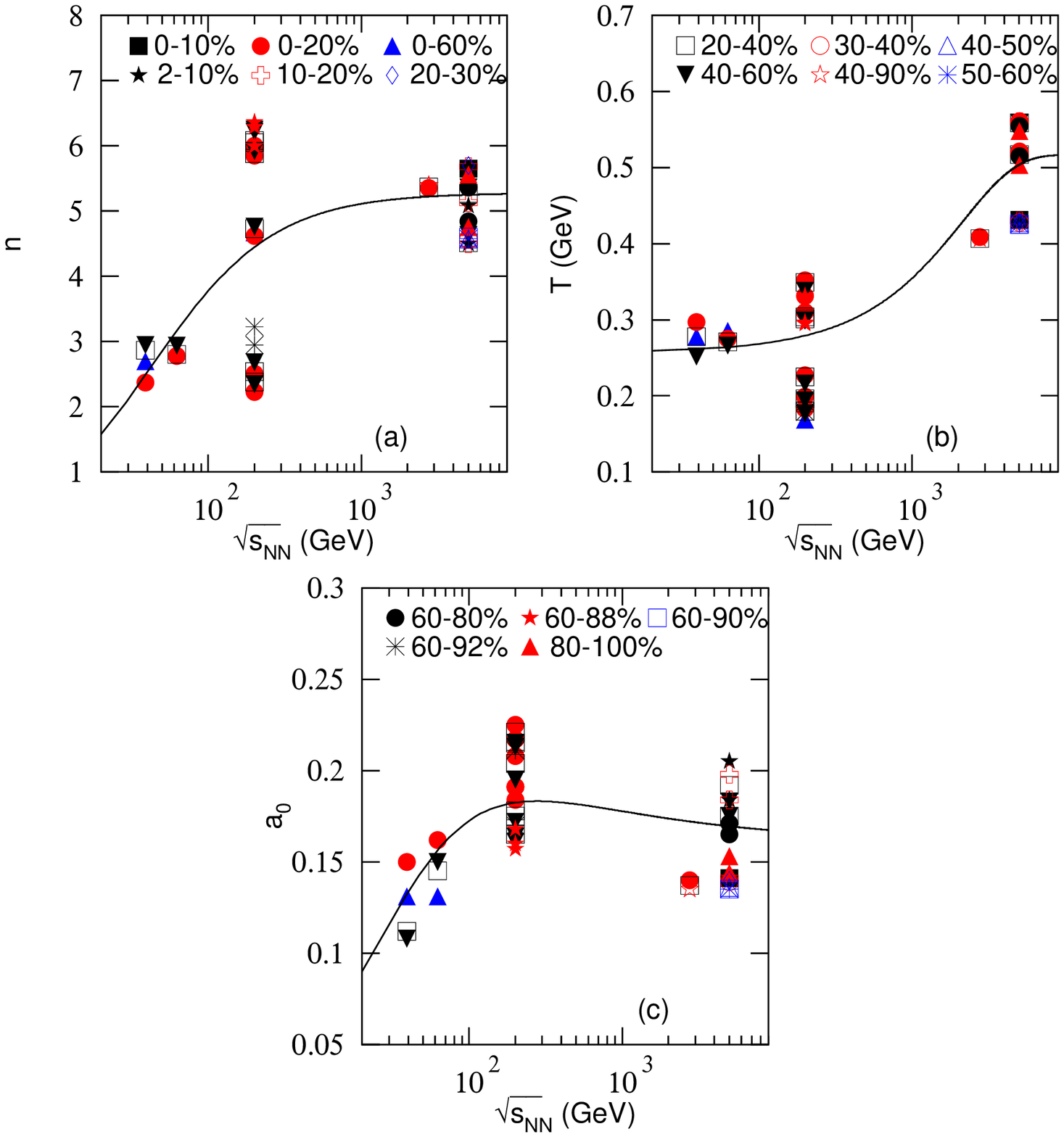}
\end{center}
\justifying\noindent {Figure 10. Dependences of (a) $n$, (b) $T$,
and (c) $a_{0}$ on $\sqrt{s_{\rm NN}}$ for different centrality
classes in Au-Au, $d$-Au, Pb-Pb, and $p$-Pb collisions at the RHIC
and LHC. The symbols represent the parameter values listed in
Tables 1--4 and obtained from Figures 1--4. The curves are the
results fitted by empirical formulas for guide the eyes.}
\end{figure*}

\subsection{Tendencies of parameters}

Due to the worse results obtained from the fit function with fixed
$a_0=1$, we give up to analyze the changing laws of the parameters
listed in the right panel in the above tables and extracted from
the dot-dashed curves in the above figures. To better analyze the
changing laws of the parameters used in the fit function with
changeable $a_0$ for different centralities, system sizes, and
energies, Figure 6 shows the dependences of the parameters (a)
$n$, (b) $T$, and (c) $a_{0}$ on centrality percentage $C$ in
Au-Au, $d$-Au, Pb-Pb, and $p$-Pb collisions at the RHIC and LHC.
The symbols represent the parameter values listed in the left
panel in Tables 1--4 and extracted from the solid curves in
Figures 1--4. One can see that with the increase of $C$ (with the
decrease of centrality from central to peripheral collisions), the
power exponent $n$ increases slightly, the effective temperature
$T$ does not change obviously, and the revised index $a_0$
decreases slightly in most cases. At the given energy (the top
RHIC or LHC energy), the values of $n$, $T$, and $a_0$ for $d$-Au
and $p$-Pb collisions are larger than those for Au-Au and Pb-Pb
collisions.

For given collision system, the tendencies in Figure 6 indicate
that the spectra of $J/\psi$ are affected slightly by the
centrality or impact parameter. The size of participant region and
the effect of cold spectator nucleons have weak influences on the
production of $J/\psi$. The $J/\psi$'s produced in the early stage
of collisions have less interactions with the hot and dense QGP
and the cold spectator nucleons. For different collision systems,
the system size dependence of parameters is caused by the
secondary cascade collisions of produced $J/\psi$ with the
subsequent nucleons in the incident path way of $d(p)$ in target
Au(Pb), where the subsequent nucleons are the remainder nucleons
after the binary nucleon-nucleon collisions in the incident path
way. The subsequent nucleons are also the cold spectator nucleons
in $d$-Au and $p$-Pb collisions. It is believed that the
subsequent nucleons have larger influence on $J/\psi$ than other
cold spectator nucleons outside the incident path way. In Au-Au
(Pb-Pb) collisions, there is no subsequent nucleon after the
binary nucleon-nucleon collisions. In most cases, the values of
$n$ are large enough, which implies that the entropy index $q$ is
close to 1 due to the fact that $n=1/(q-1)$ in the TP-Like
function. This means that the emission source of $J/\psi$ and the
collision system stay approximately at equilibrium.

Figure 7 shows the dependences of (a) $n$, (b) $T$, and (c)
$a_{0}$ on rapidity $y$ in the forward region, which are extracted
from the spectra of prompt $J/\psi$ and $J/\psi$ from $b$ in
$p$-$p$ collisions at the LHC (5.02, 7, 8, and 13 TeV). The
symbols represent the parameter values listed in the left panel in
Table 5 and obtained from the solid curves in Figure 5. One can
see that with the increase of $y$, the parameters $n$ and $T$
increase and the parameter $a_0$ decreases. The three parameters
depend on the type of $J/\psi$. For prompt $J/\psi$, $n$ is
slightly larger and $T$ and $a_0$ are significantly less than
those for $J/\psi$ from $b$.

Being the parameter which describes the excitation degree of
emission source, the effective temperature $T$ of emission source
for prompt $J/\psi$ is lower than that for $J/\psi$ from $b$. This
implies the more energy deposition, lower threshold energy, and
larger particle yields in the production process of prompt
$J/\psi$. The $J/\psi$ from $b$ is produced from the fragmentation
of open bottom hadron. Its emission source has larger $T$ and
smaller $n$ due to larger mass of bottom, which has a harder
spectrum than charm. Not only for prompt $J/\psi$ but also for
$J/\psi$ from $b$, we have used the convolution of two TP-like
functions to fit the spectra.

The situation of emission source for $J/\psi$ is different from
light particles which are produced in the later stage of
collisions. As early produced particle, $J/\psi$ at forward
rapidity means less influence from the hot and dense QGP region,
then higher degree of equilibrium and higher temperature of the
emission source in the early stage are obtained. Indeed, in the
forward rapidity region, the larger power exponent $n$ implies the
entropy index $q$ being closer to 1 and higher degree of
equilibrium of the emission source. Compared with that for
$J/\psi$ with small rapidity, the larger $T$ implies that $J/\psi$
with large rapidity loses less energy when it goes through the hot
and dense QGP region due to large flying velocity and less
interaction time. Compared to the light particles produced in the
later stage, $J/\psi$ produced in the early stage has different
properties of source.

The dependences of (a) $n$, (b) $T$, and (c) $a_{0}$ on $y$ in the
forward and backward regions, which are extracted from the spectra
of prompt $J/\psi$ and nonprompt $J/\psi$ in $p$-Pb and $p$-$p$
collisions at 5.02 TeV, are displayed in Figure 8. The symbols
represent the parameter values listed in the left panel in Tables
4 and 5 and obtained from the solid curves in Figures 4 and 5. One
can see that with the increase of $|y|$, the parameters $n$ and
$T$ increase and the parameter $a_0$ decreases. The three
parameters depend on the types of $J/\psi$ which are produced from
different sources. For prompt $J/\psi$, $n$ is slightly larger and
$T$ and $a_0$ are significantly less than those for nonprompt
$J/\psi$. These conclusions are consistent with Figure 7.

The discussions on the tendencies of parameters for Figure 7 are
suitable to those for Figure 8, if the term ``$J/\psi$ from $b$"
is replaced by ``nonprompt $J/\psi$". The conclusions from Figures
7 and 8 confirm the current knowledge of $J/\psi$ from different
sources.

The dependences of (a) $n$, (b) $T$, and (c) $a_{0}$ on energy
$\sqrt{s_{\rm NN}}$ or $\sqrt{s}$ in the forward rapidity region,
which are extracted from the spectra of prompt $J/\psi$ and
$J/\psi$ from $b$ in $p$-Pb and $p$-$p$ collisions at the LHC
(5.02, 7, 8, and 13 TeV), are displayed in Figure 9. The symbols
represent the parameter values listed in the left panel in Tables
4 and 5 and obtained from the solid curves in Figures 4 and 5. One
can see that with the increase of $\sqrt{s_{\rm NN}}$ or
$\sqrt{s}$ from 5.02 to 13 TeV, the three parameters do not show
an obvious change. The relative sizes of the parameters for
emissions of prompt $J/\psi$ and $J/\psi$ from $b$ are consistent
with those from Figures 7 and 8.

The reason for the less obvious change is that the energy range
considered in Figure 9 is very narrow, or the parameters saturate
indeed in the considered energy range. Even if there are changes
in the parameters with an increase in energy, it is hard to detect
them in the narrow energy range. The production of prompt $J/\psi$
is from approximate equilibrium due to the fact that $n$ is large
enough (and hence $q$ is close to 1). The production of $J/\psi$
from $b$ is also from approximate equilibrium, though its
production source is very different from prompt $J/\psi$.
Comparing with the production of $J/\psi$ from $b$, the production
of prompt $J/\psi$ corresponds to lower excitation degree (smaller
$T$) of emission source, which results in steeper spectrum and
smaller $a_0$.

Figure 10 gives the dependences of (a) $n$, (b) $T$, and (c)
$a_{0}$ on $\sqrt{s_{\rm NN}}$ for different centrality classes in
Au-Au, $d$-Au, Pb-Pb, and $p$-Pb collisions at the RHIC and LHC.
The symbols represent the parameter values listed in the left
panel in Tables 1--4 and obtained from the solid curves in Figures
1--4. The curves are the results fitted by empirical formulas
which will be discussed later. One can see that $n$ and $T$
increases and $a_0$ does not change obviously with the increase of
$\sqrt{s_{\rm NN}}$ when it varies from the RHIC to LHC energies.

The parameter tendencies in Figure 10 indicate that, at the LHC,
the higher collision energy results in more energy deposition in
the collisions and higher excitation degree of the emission
source. Then, more yields are produced, and quicker equilibrium is
approached, from the statistical point of view. The almost
invariant $a_0$ reveals that the spectrum shapes and hence the
production mechanisms of $J/\psi$ at the RHIC and LHC are similar.
Or, in the two energy ranges, the early produced $J/\psi$ traverse
through similar participant regions, in which the matters formed
in both the cases are hot and dense QGP. The initial energy of QGP
formation is below 39 GeV~\cite{68}.

Based on the above discussions we see that Eq. (3) is a
generalization of Eq. (2) with one more parameter, and surely the
TP-like function of Eq. (3) can fit experimental data better. As
showed in Figures 1--5 by the curves and listed in Tables 1--5 for
the $\chi^2$/ndof, we have obtained several times larger $\chi^2$
with fixed $a_0=1$ than those with changeable $a_0$ in most cases.
That is, if we do not introduce the parameter $a_0$ in Eq. (3) and
utilize directly Eq. (2) or the convolution Eq. (6) with $a_0=1$
to fit the data, worse fits will be obtained, though in some cases
the two fits are comparable each other. So, we summarize that the
introduction of $a_0$ is indeed necessary.

It seems that Tables 1--5 show that for each different set of data
we need a different set of four parameters (three free parameters
and one normalization constant), and there are no obvious
relations between 4 parameters of different sets. In fact, there
are tendencies of free parameters on centrality, rapidity, and
collision energy. Then, we may predict some results in central
collisions at other energies.

For example, the fitted curves in Figure 10 can be approximately
parameterized by (a) $n=a_1+b_1/[1+c_1(\sqrt{s_{\rm NN}})^{d_1}]$
(with $a_1=5.279$, $b_1=-5.361$, $c_1=0.018$, and $d_1=1.082$),
(b) $T=a_2+b_2\exp\{-2[(\sqrt{s_{\rm NN}}-c_2)/d_2]^{2}\}$ (with
$a_2=0.517$, $b_2=-1.832$, $c_2=-6901.400$, and $d_2=7924.553$),
and (c) $a_0=\sqrt{s_{\rm NN}}/[a_3+b_3\sqrt{s_{\rm
NN}}+c_3(\sqrt{s_{\rm NN}})^{1/2}]$ (with $a_3=210.596$,
$b_3=6.214$, and $c_3=-25.252$), respectively, where $\sqrt{s_{\rm
NN}}$ is in the units of GeV. For 0--20\% Pb-Pb collisions at 8.8
TeV, we may obtain approximately $n=5.26$, $T=0.516$ GeV, and
$a_0=0.168$ which are less centrality dependent according to
Figure 6. Because of different types of spectra being used in
experiments, the normalization constants are not uniform. We have
not given the normalization constant for 0--20\% Pb-Pb collisions.

\subsection{Further discussions}

Before summary and conclusions, we would like to point out that
the effective temperature $T$ discussed above is not the real
temperature. Because the effect of transverse flow is not
excluded, we call $T$ the effective temperature. If we exclude the
effect of transverse flow, a smaller $T$ will be obtained.
However, being an early produced particle, $J/\psi$ has less
influence from the transverse flow which appears in the later
stage of collisions. So, we think that the values of $T$ extracted
from $J/\psi$ spectra in this work can be approximately regarded
as the initial temperature $T_i$ of the collision system. Because
prompt $J/\psi$ and $J/\psi$ from $b$ are from different sources,
the initial temperatures extracted from the two $J/\psi$ spectra
may be different. This is a reflection of the multi-source picture
discussed by us.

Generally, the initial temperature $T_i$ is larger than the
chemical freeze-out temperature $T_{ch}$ ($\sim160$ MeV) which is
extracted from the ratios of different particle yields. $T_{ch}$
is larger than the kinetic freeze-out temperature $T_{kin}$ or
$T_0$ which is extracted from the transverse momentum spectra in
the last stage of collisions. The results of the present work
support this statement. Although the transverse flow affects
slightly $T_i$, even if it does not affect $T_{ch}$, it affects
obviously $T_0$. Because the transverse flow affects the
transverse momentum spectra in the last stage of collisions, it
also affects the quantities extracted from the mentioned spectra.

The revised index $a_0$ is a dimensionless quantity. However, its
introduction causes the dimension or unit of $p_{\rm T}^{a_0}$ to
be $({\rm GeV}/c)^{a_0}$. This unit should be combined with the
unit of normalization constant $C$ in Eq. (3) to result in the
unit of $f(p_{\rm T})$ to be $({\rm GeV}/c)^{-1}$. That is to say,
the unit of $C$ in Eq. (3) is $({\rm GeV}/c)^{-a_0-1}$. The values
of $a_0$ are changeable. Thus, the units of $p_{\rm T}^{a_0}$ and
$C$ are also changeable. As a result, the unit of $f(p_{\rm T})$
is always $({\rm GeV}/c)^{-1}$ which is independent of $a_0$.

In the above discussions, we have used the concept of equilibrium.
This means that lots of particles should be considered. However,
in many cases, the yields of $J/\psi$ are not too large. To
consider lots of particles, we may include lots of events which
are from the same collisions. Although the particles from
different events are not correlated, their production proceeds
under similar conditions, which allows increasing their
statistics. From the statistical point of view, the particle
productions in high energy collisions are a statistical process,
rather than a thermodynamical process, though some quantities such
as temperature can be used to connect the two processes.

In low statistics with single or few events, the concept of
temperature can be also used, if the temperature reflects the
average kinetic energies of the considered particles, even the
kinetic energy of a given particle. Anyhow, the temperature
reflects the excitation degree of emission source. A low
excitation degree of emission source corresponds to a low kinetic
energy of the considered particle. This is similar to the case in
single atomic and molecular physics, in which a cold atom or
molecule has a low kinetic energy and then corresponds to a low
temperature of emission source.

At the level of partons, we have an alternative discussion on the
temperature of emission source in the collision system. Generally,
we may think that lots of gluons and sea quarks taking part in the
collisions, though only two partons are considered in Eq. (6) to
contribute to the transverse momentum of particle. These partons
stay at equilibrium or local equilibrium state. The concept of
temperature is suitable and the transverse momentum or kinetic
energy of particle is a reflection of the temperature.

Although the tendencies of parameters have been obtained in the
above discussions, the main significance of the present work is in
the field of methodology. According to the probability density
functions (the TP-like functions) contributed by the two
contributor partons, the transverse momentum distribution of
particles is described by the convolution of two TP-like
functions. The same idea can be used in the description of the
spectra of different particles and jets.

The TP-like function is generalized from Eq. (2) which is one form
of the Tsallis
distribution~\cite{42,43,44,45,46,47,48,59,60,61,62,63}. In those
studies, although the spectra of $J/\psi$ are approximately
described by the Tsallis distribution, the present fits are better
in most cases due to smaller $\chi^2$/ndof. In addition, the
present fits are suitable for the spectra of different particles
which include leptons, mesons, and baryons~\cite{64,64a}, as well
as various jets~\cite{65} due to the introduction of $a_0$. Not
only for other studies but also for this work, $q$ is close to 1
which means the system or parton sources being approximately in
equilibrium when different particles emit. However, the partons
are not in equilibrium in some cases when jets emit.

Except the widely applications of the Tsallis distribution with
different forms~\cite{42,43,44,45,46,47,48,59,60,61,62,63}, there
are some studies on the Tsallis distribution itself. The Tsallis
distribution with $q>1$ may cover two or three Boltzmann-Gibbs,
Fermi-Dirac, or Bose-Einstein distributions~\cite{90a}. If one
associates the Tsallis distribution~\cite{46,91,92} with the
Tsallis statistics, the Tsallis distribution belongs to the
Tsallis-2 statistics~\cite{93,94,95,96}. In the case when one
associates the Tsallis distribution with the $q$-dual statistics
which has infinite terms~\cite{96}, the Tsallis distribution is
more concordant with statistical physics.

\section{Summary and Conclusions}

We have aggregated the transverse momentum spectra of $J/\psi$
produced in high energy collisions measured by several
collaborations at the RHIC, Tevatron, and LHC. Two contributor
partons, a charm quark and an anti-charm quark, are considered in
the production of $J/\psi$. The contribution of each quark to
$J/\psi$ transverse momentum is described by the TP-like function.
The convolution of two TP-like functions is successful in fitting
the $J/\psi$ spectrum. Three free parameters, the power exponent
$n$, effective temperature $T$, and revised index $a_0$ are
extracted from the spectra with different centrality classes,
rapidity intervals, and energy ranges.

With the decrease in collision centrality from central to
peripheral collisions, $n$ increases slightly, $T$ does not change
noticeably, and $a_0$ decreases a little bit in most of the cases.
The spectra of $J/\psi$ are not affected largely by the
centrality, or by the size of participant region and the effect of
cold spectator nucleons. The $J/\psi$ produced in the early stage
of collisions has less interactions with the hot and dense QGP and
the cold spectator nucleons. The emission source of $J/\psi$ and
the collision system stay approximately at equilibrium.

With the increase of rapidity, $n$ and $T$ increase and $a_0$
decreases. For prompt $J/\psi$, $n$ is slightly larger and $T$ and
$a_0$ are significantly less than those for $J/\psi$ from $b$ or
nonprompt $J/\psi$. Comparing with $J/\psi$ from $b$ or nonprompt
$J/\psi$, in the production process of prompt $J/\psi$, there are
more energy deposition, lower threshold energy, lower excitation
degree, steeper spectrum, and more particle yields. In the forward
rapidity region, the emission source approached a more
equilibrium. $J/\psi$ with large rapidity lost less energy when it
undergone through the hot and dense QGP region.

With the increase of energy, $n$ and $T$ increases and $a_0$ does
not change obviously. The higher collision energy results in more
energy deposition, higher excitation degree, more particle yields,
and quicker equilibrium. At the RHIC and LHC, the spectrum shapes
and the production mechanisms of $J/\psi$ are similar. The early
produced $J/\psi$ undergo through similar hot and dense QGP
regions with less influence from the transverse flow. The values
of $T$ extracted from $J/\psi$ spectra can be approximately
regarded as the initial temperature of the system.
\\

\section*{Acknowledgments}

The work of X.H.Z. and F.H.L. was supported by the National
Natural Science Foundation of China under Grant Nos. 12047571,
11575103, and 11947418, the Scientific and Technological
Innovation Programs of Higher Education Institutions in Shanxi
(STIP) under Grant No. 201802017, the Shanxi Provincial Natural
Science Foundation under Grant No. 201901D111043, and the Fund for
Shanxi ``1331 Project" Key Subjects Construction. The work of
K.K.O. was supported by the Ministry of Innovative Development of
Uzbekistan within the fundamental project on analysis of open data
on heavy-ion collisions at RHIC and LHC.
\\
\\
{\bf Data availability statement}

This manuscript has no associated data or the data will not be
deposited. [Authors' comment: The data used to support the
findings of this study are included within the article and are
cited at relevant places within the text as references.]
\\
\\
{\bf Compliance with ethical standards}
\\
\\
{\bf Ethical approval}

The authors declare that they are in compliance with ethical
standards regarding the content of this paper.
\\
\\
{\bf Disclosure}

The funding agencies have no role in the design of the study; in
the collection, analysis, or interpretation of the data; in the
writing of the manuscript; or in the decision to publish the
results.
\\
\\
{\bf Conflicts of interest}

The authors declare that there are no conflicts of interest
regarding the publication of this paper.

\end{document}